\documentclass{aa}

\def\et{et al.}                        

\newcommand\mconc[1]{\multicolumn{1}{c}{#1}}
\newcommand\mctwc[1]{\multicolumn{2}{c}{#1}}
\newcommand\mcthc[1]{\multicolumn{3}{c}{#1}}

\newcommand\mcfic[1]{\multicolumn{5}{c}{#1}}

\begin{document}

   \title{Absolute kinematics of radio source components in the complete S5
polar cap sample}

   \subtitle{I. First and second epoch maps at 8.4\,GHz}

   \author{E.\ Ros\inst{1}
\and J.M.\ Marcaide\inst{2}
\and J.C.\ Guirado\inst{2,3}
\and M.A.\ P\'erez-Torres\inst{4}
          }

   \offprints{E. Ros, ros@mpifr-bonn.mpg.de}

   \institute{
Max-Planck-Institut f\"ur Radioastronomie,
Auf dem H\"ugel 69, D--53121 Bonn, Germany
\and
Departament d'Astronomia i Astrof\'{\i}sica, Universitat de Val\`encia,
E-46100 Burjassot, Val\`encia, Spain
\and
Observatorio Astron\'omico, Universitat de Val\`encia,
E-46100 Burjassot, Val\`encia, Spain
\and
Istituto di Radioastronomia, Via Gobetti 101, I-40129 Bologna, Italy
             }

\date{Submitted: 20 April 2001 / Accepted: 4 July 2001}

\abstract{
We observed the thirteen extragalactic radio sources  
of the S5 polar cap sample at 8.4\,GHz with the Very Long Baseline Array, 
on 1997.93 and 1999.41. 
We present the maps from those two epochs and
briefly discuss the morphological changes experimented by
some of the radio sources in the 1.4\,yr elapsed.  
These results correspond to the first two epochs at
8.4\,GHz of a program
directed to study the absolute kinematics of the radio source 
components of the members of the sample by means of phase delay
astrometry at 8.4, 15 and 43\,GHz.
}

   \maketitle

\section{Introduction\label{sec:introduction}}

Very-long-baseline interferometry (VLBI) 
allows routine imaging of compact radio sources
with milliarcsecond resolution.
Use of phase-delay astrometric techniques in VLBI
also allows determinations of 
relative positions with nearly microarcsecond accuracy
(e.g., Marcaide \& Shapiro \cite{mar83}).  
Moreover, phase delay astrometry, unlike group-delay astrometry, 
permits a reliable subtraction of the source structure
contribution to the delay, by referring such contribution
to a reference point in the map.
Such a (fixed) reference point provides an unambiguous
identification of radio source features seen at different epochs,
crucial for the determination of the absolute kinematics
of the components
(see Charlot \et\ \cite{cha90},
Guirado \et\ \cite{gui95b}, Ros \et\ \cite{ros99}).

In the past, our group has successfully applied 
phase-delay astrometry techniques 
to a number of pairs of radio sources.
Guirado \et\ (\cite{gui95a,gui98}) studied the
pair \object{QSO\,1928+738}/\object{BL\,2007+777} 
(about 5$^\circ$ apart) at 2.3, 5, 
and 8.4\,GHz. 
Ros \et\ (\cite{ros99})
added a new source (\object{BL\,1803+784}) to this pair, 
and extended the phase-connection technique 
(Shapiro \et\ \cite{sha79}) {to separations of 6\fdg8}
from observations at 8.4\,GHz.  
P\'erez-Torres \et\ (\cite{per00}) successfully 
applied the phase-connection technique to the radio sources \object{QSO\,1150+812}/\object{BL\,1803+784} (nearly  15$^\circ$ apart), from
observations at 2.3 and 8.4\,GHz. 
{In addition, 
Ros \et\ (\cite{ros00}) and P\'erez-Torres \et\ (\cite{per00})
have demonstrated that
dual-frequency observations
are not required anymore to subtract the ionospheric delay from
the data.}
Indeed, those authors showed that it is possible to accurately 
model the ionosphere contribution from Global Positioning System data taken at
sites nearby VLBI stations.  

The astrometric results mentioned above have encouraged us {to try}
to demonstrate the feasibility of the phase-connection technique
for a complete sample of radio sources, using single-frequency VLBI 
observations, with the aim of studying the absolute kinematics
of all the sources in the sample. 
Our sample consists of the 13 sources selected by  
Eckart \et\ (\cite{eck86,eck87}) from
the  S5 survey (K\"uhr \et\ \cite{kuh81}),
{with the following selection criteria}:
a) declination $\delta\ge70^{\circ}$, galactic latitude
$|b_{\rm II}|\ge10^{\circ}$, b) flux density
$S_{\rm 5\,GHz}\ge1$\,Jy at the epoch of the survey, and c) spectral
index $\alpha_{\rm 
2.7,\,5\,GHz}\ge-0.5$ ($S\sim\nu^{+\alpha}$).  
Throughout the paper, we will refer to those 13 radio sources as 
``the complete S5 polar cap sample".

\begin{figure}
\vspace{8cm}
\includegraphics{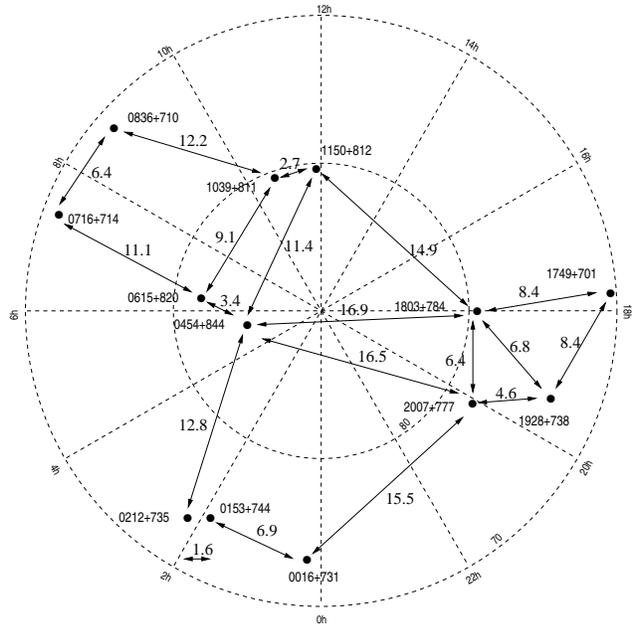}
\caption{Distribution of the complete S5 polar cap sample in the northern sky,
centered at the celestial north pole. The black dots represent the positions
of radio sources, and the angular distances among them are indicated in
arc degrees with arrows.}
\label{fig:polcap}
\end{figure}

We show in Fig.\ \ref{fig:polcap} the sky distribution of the
members of the complete S5 polar cap sample, 
{and indicate their names and relative
angular separations.}
All 13 members of this sample are at such close angular distances
that the use of an appropriate observing scheme
permits a successful phase-connection of the 
data for a 24-hour observing Very Long Baseline Array (VLBA) run. 
Provided that enough phase-delay data 
(usually several hours) are acquired for each of the 13 radio sources, 
we should be able to solve for their positions 
with accuracies better than 0.1\,milliarcseconds (mas).
{Once we precisely determine the relative positions of
all the members of the S5 polar cap sample at 8.4\,GHz for several
epochs, the registration of the maps should 
be straightforward.}

Such an astrometric analysis 
should allow us to unambiguously discern -- for each of
the 13 radio sources -- which components
are moving and which are stationary. 
Finding stationary components
in 15 sources (the 13 of our sample, 3C\,345 (Bartel \et\ \cite{bar86}),
and 1038+528\,A (Marcaide \et\ \cite{mar94})) 
would strongly support the standard model for AGNs (Blandford \& K\"onigl 
\cite{bla79}). 
On the contrary, finding that
some have no stationary components might reveal phenomena 
unthought. 
Knowing which components move and how, in an absolute sense, 
should allow us to propose the correct physical models. 

As a first step in the astrometric process, we present here the hybrid maps 
of all radio sources {at 8.4\,GHz} for the two observing epochs.
{The} astrometric results will be shown in
further publications. 

We concisely describe the {8.4\,GHz} observations in 
Sect.\ \ref{sec:observations}, 
and present the {source maps corresponding to}
the first (1997.93) and second (1999.41) epochs for all 
13 radio sources in Sect.\ \ref{sec:results}. 
{We} conclude with some summary remarks in Sect.\ \ref{sec:summary}.

%

\section{Observations\label{sec:observations}}

We observed the complete S5 polar cap sample at 8.4\,GHz on 6 December 1997 (epoch
1997.93) and on 28 May 1999 (epoch 1999.41) with the VLBA, each time 
for 24 hours. 
We used each time the recording mode 128--8--1 (1 bit sampling,
16 channels per each of the 8 IFs, with a total
of 128 channels), 
yielding a total bandwidth of 64\,MHz, 
in right circular polarization.
Data were correlated 
at the 
VLBA Array Operations Center of the National Radio 
Astronomy Observatory (NRAO) in Socorro, New Mexico, 
using a basic integration time of 4\,s. 
All radio sources were detected and provided fringes
for all baselines.  
We observed in {groups} of 3 or 4 radio sources 
in a cyclic way with duty cycles of about 6\,min.
Every scan was 78\,s long, and the small ({a few} seconds) time gap 
between different scans 
was used for slewing the VLBA antennas.  
We replaced one or two members of the group
{of observed sources} by new
members {about} every 2 hours, until all 13 members 
of the sample were observed. 
This observing scheme resulted in a time coverage for 
each single radio source of 
about 4 hours. 
We tracked the clock behavior -- of relevance for the ulterior astrometric 
analysis -- by including some scans of 
\object{BL\,0454+844} in all {groups} observed during the
first 12 hours, and of \object{BL\,2007+777} in 
all {groups} observed during the second 12 hours.
 
After having fringe-fitted the correlator output, 
we obtained correlation amplitudes 
for each source. 
We then constructed the visibility amplitudes in
a standard manner, using the antenna gain and
system temperature information provided by each station.  
We exported the data for use with the Caltech VLBI Package
(Pearson \cite{pea91}) and the difference
mapping software {\sc difmap} (Shepherd \et\ \cite{she94}).  

\begin{table*}[tbhp]
\begin{flushleft}
\caption{Map parameters for {sources of} the complete 
S5 polar cap sample.}
\label{table:results-mapping}
\[
\begin{tabular}{r@{~~}c@{$\times$}crcc@{~~}c@{$\times$}crcc@{~}c}
& \mcfic{----------- Epoch 1997.93 -----------}
& \mcfic{----------- Epoch 1999.41 -----------} 
& {\scriptsize Minimum}
\\
& \mcthc{Beam FWHM$^{\rm a}$} & & 
& \mcthc{Beam FWHM$^{\rm a}$} & & 
& {\scriptsize contour}
\\
\mconc{\bf Source}  
& \mctwc{Size} & {P.A.} & {$S_{\rm peak}$} & {$S_{\rm tot}$$^{\rm (b)}$}
& \mctwc{Size} & {P.A.} & {$S_{\rm peak}$} & {$S_{\rm tot}$$^{\rm (b)}$}
& {\scriptsize level$^{\rm c}$}
\\
& \mctwc{[mas]} & & {\scriptsize [Jy/beam]} & {\scriptsize [Jy]} 
& \mctwc{[mas]} & & {\scriptsize [Jy/beam]} & {\scriptsize [Jy]}
& {\scriptsize [mJy/beam]}
\\ \hline
{\bf \object{QSO\,0016+731}}&	0.926 & 0.700 & 60\fdg1	& 0.243 & 0.528 &	1.053 & 0.649	&   7\fdg6	& 0.173	& 0.387 & 0.7 \\
{\bf \object{QSO\,0153+744}}&	1.483 & 0.609 & 25\fdg7	& 0.141	& 0.560 &	1.413 & 0.580	& --7\fdg6	& 0.101	& 0.499 & 0.8 \\
{\bf \object{QSO\,0212+735}}&	1.545 & 0.603 & 28\fdg1	& 1.489	& 2.557 &	1.499 & 0.616	& --6\fdg3	& 1.453	& 2.640 & 1.2 \\
{\bf \object{BL\,0454+844}} &	0.670 & 0.657 & 69\fdg0	& 0.149	& 0.218 &	0.685 & 0.662	&  37\fdg9	& 0.120	& 0.183 & 0.4 \\
{\bf \object{QSO\,0615+820}}&	0.819 & 0.667 & 58\fdg3	& 0.261	& 0.557 &	0.819 & 0.678	&  19\fdg7	& 0.291	& 0.566 & 1.0 \\
{\bf \object{BL\,0716+714}} &	1.454 & 0.570 & 32\fdg8	& 0.346	& 0.377 &	1.491 & 0.567	& --6\fdg8	& 0.924	& 0.990 & 0.6 \\
{\bf \object{QSO\,0836+710}}&	1.563 & 0.591 & 20\fdg9	& 1.279	& 1.942 &	1.937 & 0.885	&--19\fdg9	& 1.149	& 1.742 & 1.0 \\
{\bf \object{QSO\,1039+811}}&	0.946 & 0.621 & 19\fdg2	& 0.706	& 0.888 &	0.958 & 0.643	&--22\fdg3	& 0.721	& 0.886 & 0.8 \\
{\bf \object{QSO\,1150+812}}&	1.085 & 0.565 &--3\fdg3	& 0.648	& 1.217 &	1.090 & 0.574	&--45\fdg4	& 0.533	& 1.191 & 1.0 \\
{\bf \object{BL\,1749+701}} &	1.825 & 0.559 & 30\fdg0	& 0.347	& 0.541 &	1.823 & 0.590	&--12\fdg6	& 0.245	& 0.416 & 1.0 \\
{\bf \object{BL\,1803+784}} &	0.868 & 0.599 & 39\fdg2	& 1.471	& 2.118 &	0.988 & 0.682	& --2\fdg2	& 1.523	& 2.267 & 1.0 \\
{\bf \object{QSO\,1928+738}}&	1.907 & 0.620 & 55\fdg0	& 1.067	& 3.163 &	1.997 & 0.622	&  10\fdg0	& 1.300	& 2.979 & 1.4 \\
{\bf \object{BL\,2007+777}} &	0.822 & 0.654 & 53\fdg8	& 0.846	& 1.305 &	0.755 & 0.675	&--23\fdg5	& 0.544	& 0.969 & 0.9 \\ \hline
\end{tabular}
\]
\begin{list}{}{
\setlength{\leftmargin}{0pt}
\setlength{\rightmargin}{0pt}
}
\item[$^{\rm a}$] The restoring beam is an elliptical Gaussian
with full-width-half-maximum (FWHM) axes $a\times b$.
{For each source, the} position angle (P.A.) 
stands for the direction of the
major {axis} (measured north through east).
\item[$^{\rm b}$] Total flux density recovered in the {hybrid}
mapping process.
\item[$^{\rm c}$] Contours in the maps of the figures shown  in sections
\ref{subsec:0016} to \ref{subsec:2007}
are the tabulated value times $(-1,1,\sqrt{3},3,3\sqrt{3},\cdots)$.
\end{list}
\end{flushleft}
\end{table*}

\begin{table*}[!p]
\begin{flushleft}
\caption{Elliptical Gaussian component model parameters for the 
radio sources
of the complete S5 polar cap sample.  
}
\renewcommand{\baselinestretch}{0.75}
\label{table:modelfit}
\begin{scriptsize}
\[
\begin{tabular}{c@{}c@{\,}r@{$\pm$}l@{\,}r@{$\pm$}l@{\,}r@{$\pm$}l@{\,}r@{$\pm$}l@{\,}r@{$\pm$}l@{\,}r@{$\pm$}l@{\,}r@{$\pm$}l@{~~}r@{$\pm$}l@{\,}r@{$\pm$}l@{\,}r@{$\pm$}l@{\,}r@{$\pm$}l@{\,}r@{$\pm$}l@{\,}r@{$\pm$}l@{\,}r@{$\pm$}l}
    &  & 
\multicolumn{12}{c}{
--------------------- 
Epoch 1997.93
--------------------- 
} &
\multicolumn{12}{c}{
---------------------
Epoch 1999.41
--------------------- 
} \\
Source  &
Comp.\   
    & \mctwc{$S$} & \mctwc{$r$} & \mctwc{$\phi$} & \mctwc{$a$} & \mctwc{$b/a$} & \mctwc{$\theta$} 
    & \mctwc{$S$} & \mctwc{$r$} & \mctwc{$\phi$} & \mctwc{$a$} & \mctwc{$b/a$} & \mctwc{$\theta$} \\
    &
    & \mctwc{[mJy]} & \mctwc{[mas]} & \mctwc{} & \mctwc{[mas]} & \mctwc{} & \mctwc{} 
    & \mctwc{[mJy]} & \mctwc{[mas]} & \mctwc{} & \mctwc{[mas]} & \mctwc{} & \mctwc{} \\ \hline
%
%
\object{QSO\,0016+731} &
XA   &  145&2  &   0.8&0.2  & $-47$&14  &  0.3&0.2 &  0.4&0.3 & $-84$&4 
     &  161&2  &   0.9&0.2  & $-46$&14  & 0.2&0.1  &  0.4&0.3 & $-89$&8 \\
&XB  &  367&3  & \mctwc{0}  & \mctwc{--}& 0.7&0.1  &  0.7&0.1 & $-83$&1 
     &  206&4  & \mctwc{0}  & \mctwc{--}& 0.9&0.1  &  0.4&0.3 & $-79$&1 \\
&XC  &   12&1  &   1.2&0.2  & $+172$&10 &  1.5&0.4 &  0.4&0.2 &  $+53$&5  
     &   18&3  &   0.9&0.2  & $+178$&14 &  1.3&0.4 &  0.4&0.2 &  $+75$&4 \\ \hline
%
%
\object{QSO\,0153+744} &
XA1 &  151&5  & \mctwc{0}  & \mctwc{--} &  0.4&0.1 &  0.3&0.2 & $+75$&6 
    &  130&2  & \mctwc{0}  & \mctwc{--} &  0.5&0.1 &  0.4&0.1 & $+47$&5  \\
&XA2&   88&30 &   0.7&0.3  & $+74$&24  &   0.4&0.3 &  0.5&0.4 & $+94$&36 
    &  56&7   &   0.7&0.2  & $+68$&17  &   0.5&0.3 &  0.4&0.4 & $+11$&25 \\
&XA3&   37&15 &   1.2&0.3  & $+91$&15  &   0.4&0.1 &  0.4&0.3 &$-156$&65 
    &   30&20 &   1.0&0.4  & $+85$&24  &   0.3&0.3 &  0.6&0.4 & $-74$&50 \\
&XA4&   16&10 &   1.6&0.5  & $+103$&18 &   0.9&0.6 &  0.4&0.3 & $-32$&38 
    &   36&24 &   1.4&0.3  & $+92$&12  &   0.9&0.3 &  0.3&0.2 & $-38$&13 \\
&XC1&    5&1  &   4.7&0.2  & $+105$&2  &   1.5&0.5 &  0.4&0.3 & $-55$&29 
    &   6&1   &   4.3&0.2  & $+102$&3  &   1.0&0.2 &  0.7&0.3 & $+61$&50  \\
&XC2&    6&1  &   5.8&0.2  & $+125$&2  &   1.0&0.5 &  0.4&0.3 & $-75$&27 
    &   8&1   &   6.1&0.4  & $+115$&4  &   1.9&0.4 &  0.5&0.2 & $+20$&12  \\
&XD &   18&1  &   8.2&0.2  & $+129$&1  &   1.6&0.5 &  0.6&0.3 & $+31$&5 
    &   16&1  &   8.2&0.4  & $+128$&3  &   1.1&0.1 &  1.0&0.1 & $+45$&50  \\
&XE &   18&2  &   9.8&0.4  & $+141$&2  &   1.8&0.3 &  0.4&0.1 & $+46$&5 
    &   14&1  &   9.7&0.2  & $+140$&1  &   1.5&0.1 &  0.3&0.2 & $+57$&12  \\
&XB1&   27&15 &   9.2&0.6  & $+158$&4  &   1.6&0.2 &  0.8&0.3 & $+41$&50 
    &   28&20 &   9.3&0.2  & $+157$&1  &   1.5&0.5 &  0.6&0.1 & $+32$&30  \\
&XB2&   72&30 &  10.1&0.3  & $+155$&2  &   1.2&0.3 &  0.7&0.3 & $+86$&30 
    &   56&1  &   10.2&0.2 & $+156$&1  &   1.1&0.1 &  0.6&0.1 & $+49$&7   \\
&XB3&   65&30 &  10.4&0.4  & $+154$&2  &   1.0&0.3 &  0.7&0.2 & $+82$&34 
    &   93&20 &   10.6&0.2 & $+153$&1  &   1.1&0.1 &  0.7&0.3 &$+106$&30 \\
&XB4&   53&15 &  11.2&0.2  & $+152$&1  &   0.7&0.1 &  0.7&0.2 &$+102$&12 
    &   30&20 &   11.1&0.2 & $+152$&1  &   0.9&0.3 &  0.4&0.1 &  $+7$&30 \\ \hline
\object{QSO\,0212+735} &
XA &  353&40  & 0.5&0.2  & $-55$&23  & 0.1&0.1 &  0.4&0.3 & $-5$&30
    & 713&97  & 0.6&0.2  & $-56$&19  & 0.2&0.1 &  0.4&0.2 & $-68$&16 \\
&XB & 1847&45 & \mctwc{0}& \mctwc{--}& 0.5&0.1 &  0.7&0.1 & $-51$&2 
    & 1649&97 & \mctwc{0}& \mctwc{--}& 0.5&0.1 &  0.8&0.1 & $-77$&8 \\
&XC &  113&7  & 1.1&0.2  & $+107$&11 & 1.2&0.5 &  0.5&0.1 & $-125$&4
    &  117&6  & 1.1&0.2  &  $+96$&10 & 1.5&0.1 &  0.6&0.1 & $-131$&7 \\
&XD &  122&3  & 2.4&0.2  & $+102$&5  & 1.0&0.1 &  0.5&0.1 & $-54$&2
    &  77&4   & 2.4&0.2  & $+103$&5  & 0.7&0.1 &  0.5&0.1 & $-76$&8  \\
&XE &  11&1   & 3.9&0.2  & $+112$&3  & 0.7&0.3 &  0.4&0.3 & $-41$&21
    &  19&2   & 4.0&0.2  & $+102$&3  & 2.0&0.2 &  0.5&0.1 & $-119$&7  \\
&XF &  39&1   & 6.4&0.2  & $+103$&2  & 2.9&0.1 &  0.4&0.3 & $-67$&4
    &  17&1   & 6.3&0.2  & $+103$&2  & 2.0&0.5 &  0.4&0.2 & $-73$&4  \\
&XG &  87&1   & 13.8&0.2 &  $+93$&1  & 2.5&0.1 &  0.6&0.4 & $-49$&2
    &  72&1   & 13.7&0.2 &  $+92$&1  & 2.3&0.1 &  0.6&0.1 & $-51$&2  \\
\hline
%
\object{BL\,0454+844} &
XA   &  164&20 & \mctwc{0}&  \mctwc{--}& 0.3&0.1 &  0.7&0.1 &  $-9$&4 
     &  139&7  & \mctwc{0}&  \mctwc{--}& 0.3&0.1 &  0.7&0.1 &  $-7$&4    \\
&XB  &  37&2   & 0.6&0.2 & $+137$&20 &   0.6&0.1 &  0.2&0.2 &  $-20$&6 
     &  22&10  & 0.6&0.3 & $+139$&29 &   0.9&0.2 &  0.4&0.3 &  $-28$&8    \\
&XC  &  25&2   & 1.4&0.2 & $+167$&7  &   1.4&0.1 &  0.6&0.1 &  $-4$&4 
     &  28&3   & 1.5&0.2 & $+171$&7  &   1.6&0.1 &  0.5&0.1 &  $+15$&3    \\ \hline
%
\object{QSO\,0615+820} &
XA1  &  314&19 & \mctwc{0}& \mctwc{--} & 0.6&0.1 & 0.4&0.1 & $+31$&1 
     &  329&19 & \mctwc{0}& \mctwc{--} & 0.5&0.1 & 0.5&0.1 & $+14$&2 \\
&
XA2  &  162&19 & 0.6&0.2 & $-17$&19  &   0.7&0.1 & 0.5&0.1 & $+110$&3
     &  104&18 & 0.7&0.2 & $-60$&18  &   0.8&0.1 & 0.4&0.3 & $+80$&6 \\
&
XA3  &   80&3  & 1.0&0.2 & $-98$&12  &   1.0&0.1 & 0.3&0.3 & $-178$&2 
     &  133&4  & 0.8&0.2 & $-137$&15 &   1.2&0.1 & 0.2&0.1 & $-93$&1 \\ \hline
%
%
\object{BL\,0716+714} &
XA   &  346&16 & \mctwc{0}& \mctwc{--}& 0.2&0.1 & 0.4&0.3 &  $+18$&25
     &  933&4  & \mctwc{0}& \mctwc{--}& 0.2&0.1 & 0.2&0.1 &  $+14$&3 \\
&XB  &   18&17 & 0.8&0.4 & 11&30 & 0.5&0.3   & 0.4&0.3 &  $-57$&114
     &   47&4  & 1.0&0.2 & 12&11 & 0.7&0.1   & 0.4&0.3 & $+20$&14  \\ 
&XC  &   14&8  & 1.8&0.3 & 12&10 & 0.9&0.6   & 0.4&0.3 &  $+39$&20
     &   10&1  & 3.3&0.2 & 15&4  & 1.1&0.2   & 0.8&0.5 & $+40$&60  \\ \hline
%
\object{QSO\,0836+710} &
XA   & 1271&26 &  \mctwc{0}& \mctwc{--} & 0.1&0.1 & 0.6&0.4 &  $+89$&37 
     & 1097&21 &  \mctwc{0}& \mctwc{--} & 0.4&0.1 & 0.3&0.2 &  $+32$&1 \\
&XB  &  164&32 &  1.0&0.2 &  $-139$&11 & 1.2&0.2  & 0.3&0.1 &  $+26$&6
     &  216&21 &  0.5&0.2 &  $-130$&24 & 0.9&0.1  & 0.2&0.2 &  $+38$&2 \\
&XC  &  243&10 &  2.7&0.2 &  $-142$&4  & 1.1&0.1  & 0.4&0.1 &  $+21$&1
     &  208&1  &  2.6&0.2 &  $-144$&4  & 1.0&0.1  & 0.4&0.1 &  $-144$&1\\
&XD  &   27&2  &  4.9&0.2 &  $-149$&2  & 2.0&0.5  & 0.4&0.1 &  $+46$&4
     &   61&2  &  5.4&0.2 &  $-144$&2  & 3.8&0.1  & 0.5&0.3 &  $+33$&2  \\
&XE  &   39&1  &  7.8&0.2 &  $-140$&1  & 2.4&0.1  & 0.4&0.3 &  $+15$&1
     &   26&1  & 8.4&0.2  &  $-141$&1  & 2.0&0.1  & 0.3&0.2 &   $+8$&2 \\
&XF  &  146&1  & 11.8&0.2 &  $-148$&1  & 3.2&0.6  & 0.4&0.3 &   $+3$&1
     &  121&1  & 12.1&0.2 &  $-148$&1  & 3.1&0.1  & 0.4&0.3 &   $-1$&1  \\ \hline
%
%
\object{QSO\,1039+811} &
XA    & 737&10 & \mctwc{0}&  \mctwc{--}& 0.2&0.1 & 0.7&0.1 & $-80$&9 
      & 720&18 & \mctwc{0}&  \mctwc{--}& 0.2&0.1 & 0.4&0.1 & $-66$&2 \\
&XB   &  77&4  & 0.9&0.2 & $-63$&13 &  0.6&0.2 & 0.3&0.1 & $-71$&4 
      &  79&11 & 0.4&0.2 & $-58$&32 &  1.0&0.2 & 0.3&0.1 & $+62$&10 \\
&XC   &  46&12 & 2.1&0.2 & $-65$&6  &  0.9&0.2 & 0.7&0.1 & $-61$&10 
      &  47&9  & 1.8&0.2 & $-63$&6  &  1.3&0.3 & 0.3&0.1 & $-58$&5 \\
&XD   &  11&10 & 3.1&0.6 & $-72$&12 &  1.3&0.7 & 0.6&0.3 & $-60$&46 
      &  24&8  & 2.5&0.2 & $-71$&5  &  1.1&0.2 & 0.6&0.2 & $+63$&16 \\
&XE   &  6&2   & 5.3&0.3 & $-74$&3  &  2.6&0.4 & 0.4&0.2 & $-153$&7 
      &  10&1  & 4.8&0.2 & $-75$&2  &  1.7&0.2 & 0.8&0.2 & $-136$&36\\
&XF   &  9&1   & 7.7&0.4 & $-79$&3  &  2.2&0.4 & 0.5&0.2 & $-58$&14 
      &  8&1   & 7.4&0.2 & $-79$&2 &  1.2&0.1 & 0.4&0.3 & $-77$&8  \\ \hline
%
%
\object{QSO\,1150+812} &
XA  &  608&25 & \mctwc{0} & \mctwc{--} & 0.2&0.1 & 0.4&0.2 & $+65$&14
    &  594&17 & \mctwc{0} & \mctwc{--} & 0.3&0.1 & 0.4&0.3 & $+51$&2 \\
&XB &  226&23 & 0.5&0.2 & $+214$&17 &   0.3&0.1 & 0.4&0.3 & $+138$&1
    &  185&17 & 0.5&0.2 & $+214$&21 &   0.6&0.1 & 0.3&0.1 & $+175$&3 \\
&XC &  272&10 & 1.7&0.2 & $+182$&6  &   1.2&0.1 & 0.4&0.1 & $+12$&1
    &  258&11 & 1.8&0.2 & $+177$&6  &   1.3&0.1 & 0.6&0.1 & $-8$&2  \\
&XD &  68&11  & 2.2&0.2 & $+168$&3  &   1.0&0.1 & 0.4&0.2 & $-130$&4
    &  111&13 & 2.3&0.2 & $+176$&5  &   0.9&0.1 & 0.4&0.1 & $-72$&2 \\
&XE  &   41&1 & 4.4&0.2 & $+162$&3  &   2.5&0.1 & 0.4&0.1 & $+148$&3 
    &  45&2   & 4.4&0.2 & $+161$&3  &   2.6&0.1 & 0.4&0.1 & $+160$&1 \\ \hline
%
\object{BL\,1749+701} &
XA  & 347&12 & \mctwc{0}& \mctwc{--}& 0.2&0.1 &  0.4&0.3 & $-45$&11
    & 232&8  & \mctwc{0}& \mctwc{--}& 0.2&0.1 &  0.5&0.3 & $+133$&30  \\
&XB &  62&10 & 0.5&0.2 & $-56$&26 & 0.3&0.1 &  0.4&0.3 &  $+67$&45
    &  27&7  & 0.4&0.2 & $-53$&30 & 0.3&0.2 &  0.4&0.3 &  $+97$&60  \\
&XC &  40&7  & 1.2&0.2 & $-73$&9  & 0.5&0.1 &  0.4&0.3 &  $+89$&18
    &  57&11 & 1.2&0.2 & $-68$&10 & 0.9&0.1 &  0.9&0.1 &  $+99$&45  \\
&XD &  30&9  & 2.3&0.2 & $-64$&5  & 0.5&0.1 &  0.4&0.3 &  $-13$&26
    &  50&14 & 2.3&0.2 & $-70$&5  & 0.9&0.2 &  0.8&0.1 &  $+123$&40 \\
&XE &  53&8  & 3.2&0.2 & $-56$&4  & 1.7&0.1 &  0.7&0.4 &  $+69$&6
    &  39&9  & 3.7&0.2 & $-51$&3  & 1.8&0.1 &  0.6&0.2 &  $+62$&7   \\
&XF &  14&4  & 4.6&0.3 & $-47$&2  & 1.5&0.3 &  0.6&0.2 &  $+57$&18
    &   8&4  & 5.7&0.2 & $-52$&2  & 2.0&1.0 &  0.5&0.4 &  $+57$&18  \\ \hline
%
%
\object{BL\,1803+784} &
XA   & 1493&1 & \mctwc{0}& \mctwc{--}& 0.2&0.1 & 0.8&0.1 & $+119$&1 
     & 1517&1 & \mctwc{0}& \mctwc{--}& 0.2&0.1 & 0.6&0.1 & $+107$&1  \\
&XB  &  231&1  & 0.6&0.2 & $-86$&19 & 0.7&0.1  & 0.6&0.1 & $+102$&1
      & 374&1  & 0.5&0.2 & $-81$&21 & 0.5&0.1  & 0.7&0.1 & $+117$&1  \\
&XC  &  274&1  & 1.4&0.2 & $-97$&8  & 0.6&0.1  & 0.7&0.1 & $+41$&1 
      & 192&1  & 1.4&0.2 & $-93$&8  & 0.5&0.1  & 0.3&0.1 & $+22$&1   \\
&XD  &   36&1  & 2.0&0.2 & $-98$&5  & 0.5&0.1  & 0.2&0.1 & $-12$&4
      & 58&1   & 1.9&0.2 & $-94$&6  & 0.6&0.1  & 0.2&0.1 & $+3$&2    \\
&XE  &   24&1  & 2.7&0.2 & $-97$&4  & 1.4&0.1  & 0.4&0.1 & $+1$&3
      &  30&1  & 2.7&0.2 & $-93$&4  & 1.9&0.1  & 0.4&0.1 & $+126$&2  \\
&XF  &   18&1  & 3.9&0.2 & $-89$&3  & 1.1&0.1  & 0.9&0.1 & $-30$&16
      &  30&1  & 4.0&0.2 & $-97$&3  & 2.4&0.1  & 0.4&0.1 & $+111$&1  \\
&XG  &   19&1  & 6.3&0.2 & $-96$&3  & 2.8&0.2  & 0.5&0.1 & $+31$&4
      &  20&1  &  7.0&0.2& $-97$&2  & 2.9&0.2  & 0.6&0.1 & $+115$&5 \\
&XH  &   25&1  & 8.9&0.2 & $-95$&1  & 4.1&0.2  & 0.5&0.1 & $+99$&4
      &  17&1  & 9.4&0.2 & $-97$&1  & 3.6&0.3  & 0.6&0.1 & $+84$&5 \\ \hline
%
%
\object{QSO\,1928+738} &
XA   &  819&7  &  0.8&0.2  & $-29$&14 &   0.3&0.1 &  0.6&0.1 &  $-11$&1
     &  917&14 &  0.5&0.2  & $-25$&5  &   0.2&0.1 &  0.4&0.3 &  $-17$&3 \\
&XB  & 1267&6  & \mctwc{0}& \mctwc{--}&   0.4&0.1 &  0.5&0.1 &  $-17$&3
     &  587&8  & \mctwc{0}& \mctwc{--}&   0.2&0.1 &  0.4&0.3 &  $+120$&52 \\
&XC  &  470&7  &  1.4&0.2  & $+173$&8  &   0.7&0.1 &  0.7&0.1 &  $+50$&1
     & 630&4   &  0.8&0.2  & $+151$&15 &   0.3&0.1 &  0.4&0.3 &  $+82$&5 \\
&XD  &  192&10 &  2.0&0.2  & $+178$&6  &   0.8&0.1 &  0.4&0.2 &  $+131$&2
     &  332&2  &  2.2&0.2  & $+165$&6  &   1.0&0.1 &  0.7&0.1 &  $-9$&2  \\
&XE  &   76&8  &  3.0&0.5  & $+162$&10 &   0.7&0.1 &  0.4&0.3 &  $+4$&2
     &  185&5  &  2.7&0.2  & $+175$&5  &   1.1&0.1 &  0.4&0.3 &  $+163$&2         \\
&XF  &   37&10 &  4.1&0.2  & $+164$&3  &   0.9&0.6 &  0.6&0.1 &  $-1$&11
     &  43&2   &  3.6&0.2  & $+160$&3  &   0.6&0.1 &  0.4&0.3 &  $+64$&9 \\
&XG  &  15&7   &  5.1&0.2  & $+168$&2  &   1.2&0.1 &  0.4&0.2 &  $+82$&14
     &   54&1  &  5.1&0.2  & $+165$&2  &   1.8&0.1 &  0.4&0.1 &  $-12$&2 \\
&XH  &   86&9  &  8.7&0.2  & $+166$&1  &   1.2&0.2 &  0.6&0.2 &  $+44$&2
     &   61&3  &  9.8&0.2  & $+165$&1  &   1.6&0.1 &  0.8&0.1 &  $-4$&6  \\
&XI1$^{\rm a}$  
     &   36&10 & 10.3&0.2  & $+168$&1  &   2.1&0.1 &  0.4&0.1 &  $+49$&2
     &   47&1  & 11.4&0.2  & $+167$&1  &   1.8&0.1 &  0.4&0.1 &  $-66$&3 \\
&XI2$^{\rm a}$  
     &   46&10 & 11.7&0.2  & $+167$&1  &   1.8&0.1 &  0.6&0.1 &  $+93$&3
     & \mctwc{} & \mctwc{}& \mctwc{}& \mctwc{}& \mctwc{}& \mctwc{} \\
&XJ  &   23&9  & 14.1&0.2  & $+169$&1  &   3.5&0.3 &  0.2&0.1 &  $+78$&2
     &   32&3  & 13.2&0.3  & $+168$&1  &   3.8&0.4 &  0.2&0.1 &  $-24$&2 \\
&XK  &   68&15 & 16.3&0.2  & $+175$&1  &   4.2&0.2 &  0.7&0.1 &  $+42$&5
     &   43&2  & 16.9&0.2  & $+175$&1  &   4.7&0.3 &  0.5&0.1 &  $+29$&3 \\
&XL  &  32&10  & 20.3&0.2  & $+174$&1  &   2.6&0.2 &  0.5&0.1 &  $+52$&3
     &  29&1   & 21.1&0.2  & $+176$&1  &   2.9&0.2 &  0.6&0.1 &  $+34$&6 \\ \hline
%
\object{BL\,2007+777} &
XA   & 639&2   & \mctwc{0}& \mctwc{--}&   0.2&0.1 &  0.5&0.2 &  $-87$&1
     & 588&7   & \mctwc{0}& \mctwc{--}&   0.4&0.1 &  0.2&0.1 &  $-86$&1 \\
&XB  &  500&2  & 0.5&0.2  & $-82$&18 &   0.3&0.1 &  0.2&0.2 &  $-80$&1
     &  234&7  & 0.6&0.2  & $-88$&20 &   0.2&0.1 &  0.3&0.3 &  $+19$&7  \\
&XC  &  72&5   & 1.2&0.2  & $-102$&7 &   0.6&0.1 &  0.4&0.1 &  $-130$&3
     &  70&5   & 1.4&0.2  & $-88$&8  &   0.4&0.1 &  0.7&0.2 &  $+6$&6   \\
&XD  &   58&4  & 1.6&0.2  &  $-98$&7 &   0.7&0.2 &  0.3&0.1 &  $-77$&1
     &   41&5  & 1.7&0.2  & $-86$&7  &   0.4&0.1 &  0.2&0.1 &  $+56$&4  \\
&XE  &   23&1  & 4.7&0.2  &  $-98$&2 &   1.9&0.1 &  0.6&0.1 &  $+52$&3 
     &   8&1   & 4.8&0.2  & $-103$&2 &   2.0&1.0 &  0.4&0.3 &  $+80$&3   \\
&XF  &   15&1  & 6.7&0.2  & $-93$&2  &   1.6&0.1 &  0.5&0.1 &  $-106$&3
     &   29&1  & 6.7&0.2  & $-93$&2  &   1.2&0.1 &  0.7&0.1 &  $-113$&3 \\ \hline
\end{tabular}
\]
\begin{list}{}{
\setlength{\leftmargin}{0pt}
\setlength{\rightmargin}{0pt}
}
\renewcommand{\baselinestretch}{0.8}
\item[Note:]
For each source whose name is on the first column, the other columns
correspond to the
component designation (as indicated in Figs.\ \ref{fig:map0016} to
\ref{fig:map2007}), 
flux density, separation and position
angle relative to the strongest component, major axis, axis ratio,
and position angle of the major axis.
{The uncertainties give have been estimated from
the exploration}
of the parameter space 
($S, r, \Phi, a, b/a$, and $\theta$), and from empirical
considerations (e.g., the errors of the component positions 
are typically smaller than a tenth of a beam width,
but to be conservative we have assumed a magnitude of a fifth
of beam width for them).
\item[
$^{\rm a}$]
This component is model fitted as double only in the
first epoch.  In the second epoch the model fit algorithm merged both components
into a single one.
\renewcommand{\baselinestretch}{1.0}
\end{list}
\end{scriptsize}
\end{flushleft}
\end{table*}

\section{Imaging results\label{sec:results}}

We obtained
images of all 13 radio sources using standard hybrid mapping techniques.  
We found only minor calibration problems for data from the North Liberty
antenna at epoch 1997.93; 
we {had to multiply gain by} a factor
2.03 -- for all IF channels -- to get consistent results with 
the other nine antennas.  
We show the maps obtained in 
Figs.\ \ref{fig:map0016} through \ref{fig:map2007}. 
In all figures, east is left, and north is up.
{We list the main
parameters of those maps in Table~\ref{table:results-mapping}.}
We use a Hubble constant 
$H_0$=$65\,h$\,km\,s$^{-1}$\,Mpc$^{-1}$ and a deceleration parameter
$q_0$=$0.5$,  which results in the linear scales (10\,pc) plotted on the 
bottom right corner of each image.

\subsection{\object{QSO\,0016+731}\label{subsec:0016}}

The \object{QSO\,0016+731} has an optical magnitude $V$=19.0 and
a redshift $z$=1.781 (Stickel \& K\"uhr
\cite{sti96}).
Eckart \et\ (\cite{eck87})
reported 
that 
almost all of its flux density originates
in an area smaller than 3\,mas
and
that it does not have extended structure with
flux density larger than 0.1\,Jy at
5\,GHz.  
Pearson \& Readhead (\cite{pea88}) confirmed this result.
This source experienced a decrease of the flux density
through the nineties, 
from $\sim$1.5\,Jy for the period 1990 - 1995 to $\sim$0.4\,Jy
for the second half of the decade.

\begin{figure}[htbp]
\vspace{227pt}
\includegraphics{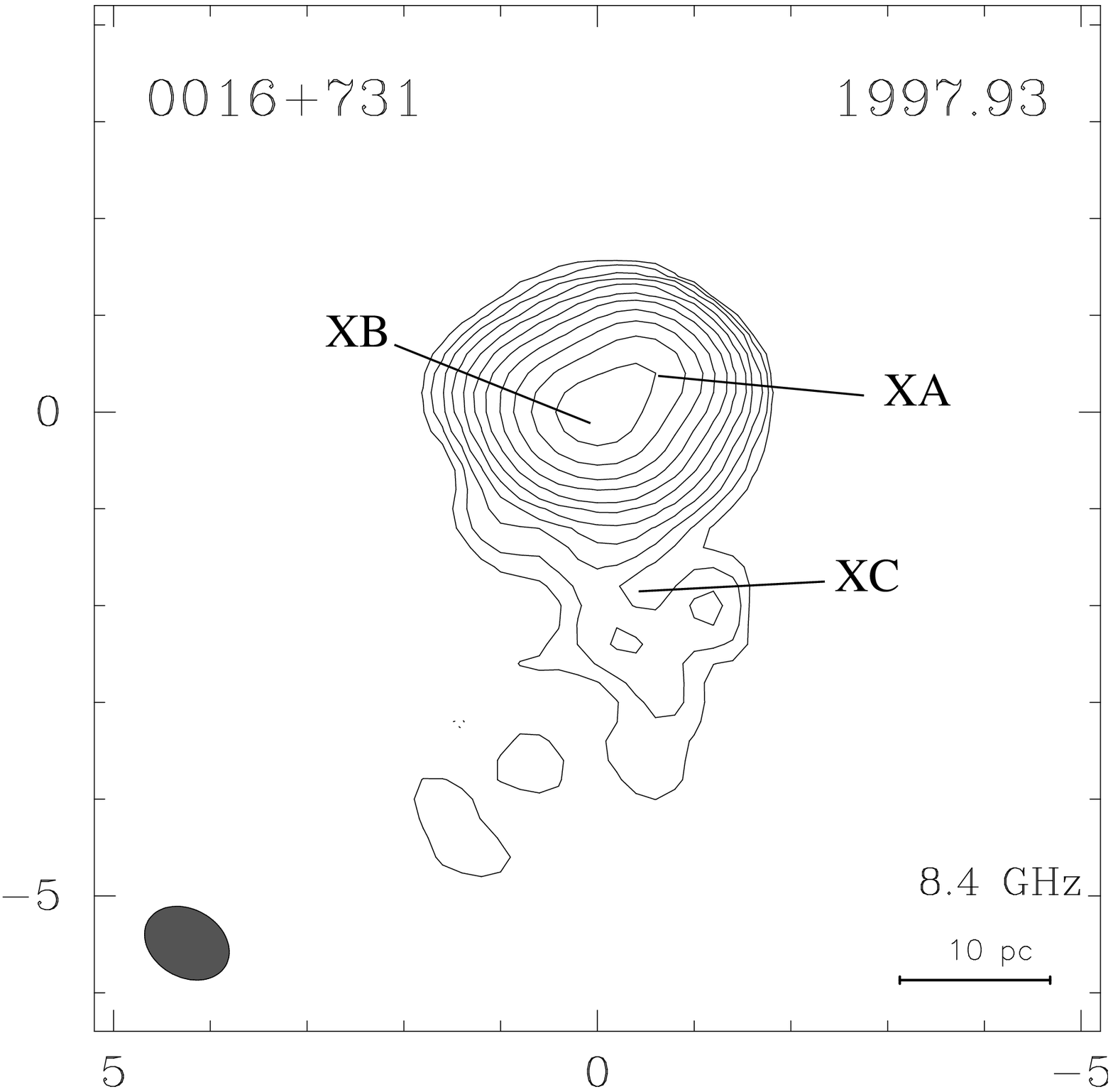}
\vspace{227pt}
\includegraphics{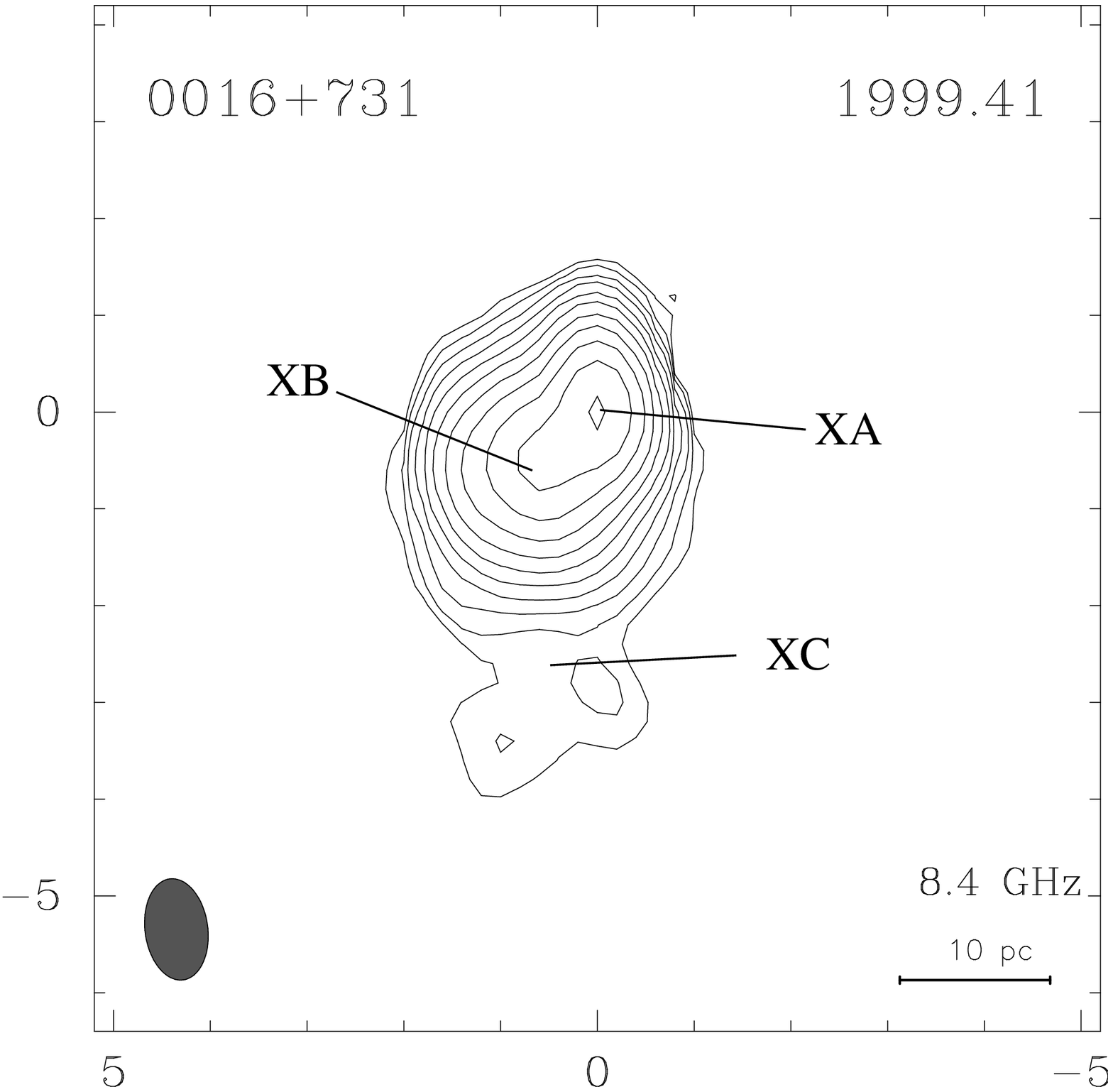}
\caption{VLBA images of \object{QSO\,0016+731} from observations on
6 December 1997 (1997.93) and 28 May 1999 (1999.41).  Axes are
relative $\alpha$ and $\delta$ in mas.  
See Table 1 for contour levels, beam sizes (bottom left in the maps) 
and peak flux densities.
{See Table~\ref{table:modelfit} for component parametrization.}}
\label{fig:map0016}
\end{figure}

Our maps (Fig.\ \ref{fig:map0016})
show a double structure along P.A.$\sim$130\degr at both epochs.  
However, an alteration in the morphology of the source is clearly seen 
between the first and the second epoch. 
The total mapped flux density decreased from 528\,mJy
to 387\,mJy ($\sim$27\,\%). 
This decrease in the total flux {density} of the source is 
due to a strong decrease in the flux of the 
westernmost component (XB, {the brightest
feature} in Fig.\ \ref{fig:map0016}).
To quantify those changes, 
we used the tasks {\sc modelfit} and {\sc modfit} of 
the Caltech package (Pearson \cite{pea91}) to
fit the visibility data to a model of elliptical components with
Gaussian brightness profiles.  
{The models are presented in Table~\ref{table:modelfit}.}
The modeling provides reasonable fits 
with three components: a compact one
(XA), a second, more elongated one (XB)
west of XA, and a third, weaker one (XC), 
the extended emission south of XA. 
The changes in the maps are likely associated to component XB,
{whose flux density decreased
from} 367\,mJy to 205\,mJy (45\%), 
while {the emission of component} XA remained more stable.

%

\subsection{\object{QSO\,0153+744}\label{subsec:0153}}

The \object{QSO\,0153+744}, has $V$=16.0 and $z$=2.338 (Stickel \& K\"uhr
\cite{sti96}).
At kiloparsec-scales, this source
does not display large structures (Hummel et al.\ (\cite{hum97}).
At pc-scales, \object{QSO\,0153+744} is a compact double radio source embedded 
in a halo, with a separation of 10\,mas between 
components (Eckart \et\ \cite{eck87}; Hummel \et\ \cite{hum88};
Pearson \& Readhead \cite{pea88}).  
{The} jet, as seen in projection, changes its direction
by a full 180\degr\ between component A and the bright
secondary component B, located about 10\,mas from the core
at P.A.\ $\sim 150$\degr.  
No component motion has been detected.
The source flux 
density at 8.4\,GHz  has been roughly stable at $\sim$0.8\,Jy,  
from 1996 to 1999.

%
\begin{figure}[htbp]
\vspace{227pt}
\includegraphics{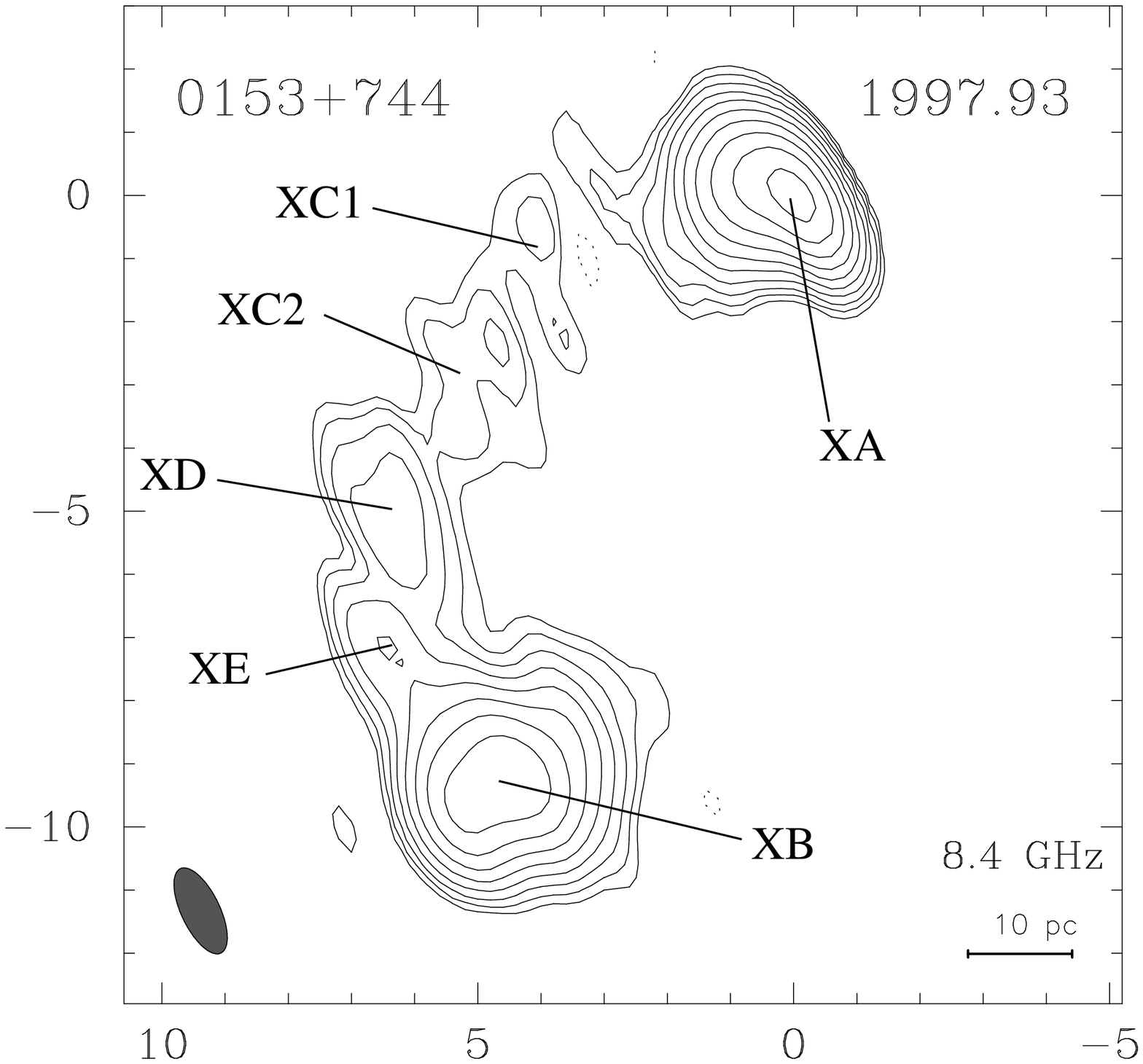}
\vspace{227pt}
\includegraphics{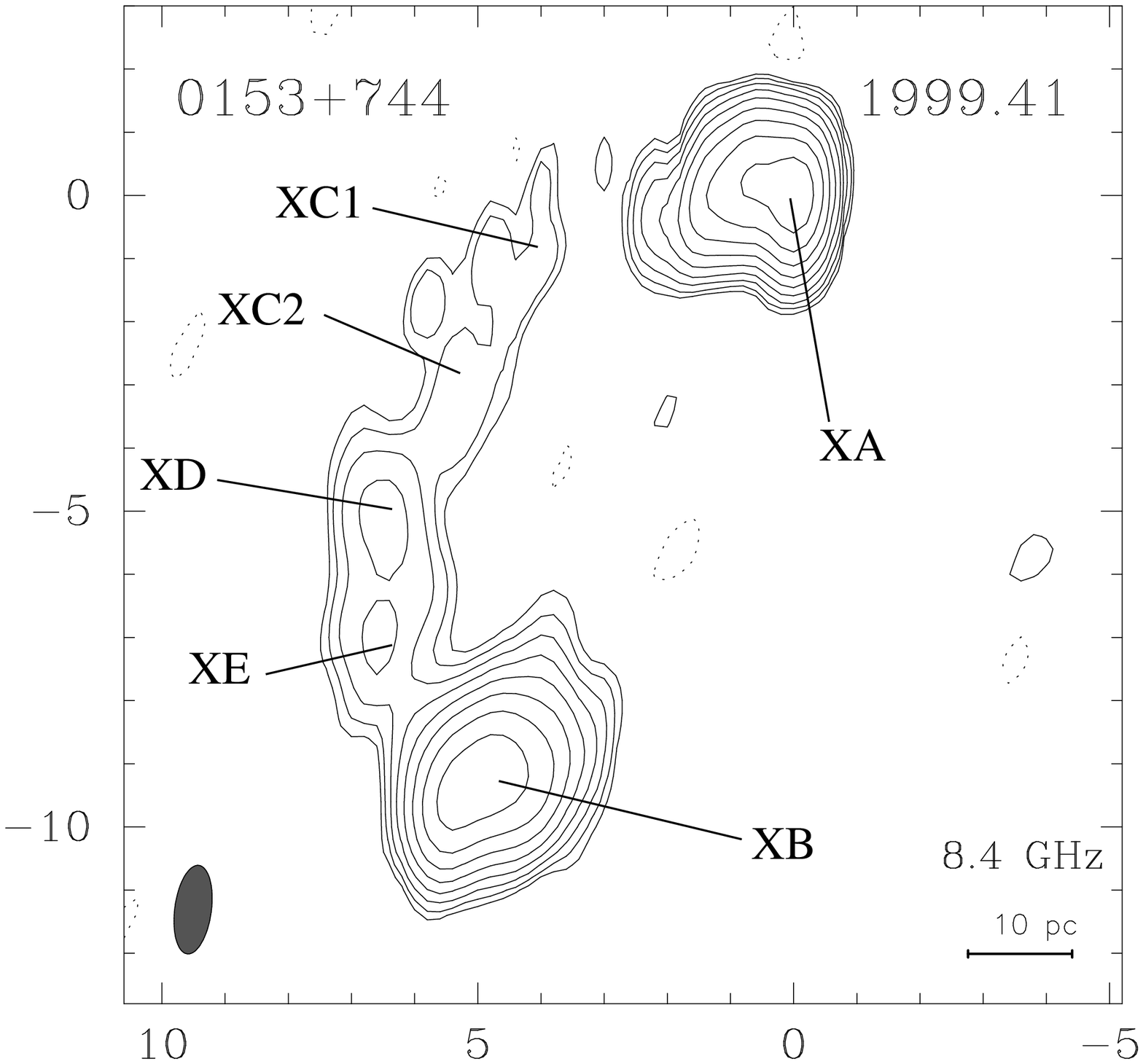}
\caption{VLBA images of \object{QSO\,0153+744} from observations
on 6 December 1997 (1997.93) and 28 May 1999 (1999.41).  Axes are
relative $\alpha$ and $\delta$ in mas. 
See Table 1 for contour levels, 
synthesized beam size (bottom left in the figure), and
peak flux densities.
{See Table~\ref{table:modelfit} for component parametrization.}}
\label{fig:map0153}
\end{figure}

Fig.~\ref{fig:map0153} shows the two {components reported}
by Eckart \et\ (\cite{eck87})
(XA and XB in the figure), and emission in an arc
joining the two components.
Both XA and XB are well modeled at each epoch by four 
Gaussian individual
components (Table~\ref{table:modelfit}).
The complex structure of this radio source 
does not show dramatic changes 
over the years, although some slight changes are readily visible.  
The overall flux density
has been decreasing slowly in the second half of the nineties
(Peng \et\ \cite{pen00}).
A careful astrometric test of the registration of the maps (Ros \et, 
in preparation),
should {shed} more light on eventual proper motions in the
components of this source.

\subsection{\object{QSO\,0212+735}\label{subsec:0212}}

The \object{QSO\,0212+735} has $V$=19.0 and $z$=2.367 (Stickel \& K\"uhr
\cite{sti96}).  This source
does not show extended emission at kpc scales in VLA images obtained 
at 1.49\,GHz
(Antonucci \et\ \cite{ant86}).  Eckart \et\ (\cite{eck87}) reported a core-jet
structure with a jet extending 12.5\,mas away from the core in 
{P.A.\,91\degr}
at 1.7\,GHz.  Pearson \& Readhead (\cite{pea88}) reported a very similar 
structure to that found by Eckart \et\ (\cite{eck87}).  
More recently, 
Fey \et\ (\cite{fey96}) reported maps at 2.3\,GHz
that show an {elongated} jet with a strong
component at 13\,mas ({P.A.\,91\degr}) of the core and a fainter component
at 41\,mas ({P.A.\,95$^\circ$}) from the core.  
Their 8.4\,GHz map shows extended emission up to 
14\,mas, in {P.A.\,42\degr}.

\begin{figure}[htbp]
\vspace{228pt}
\includegraphics{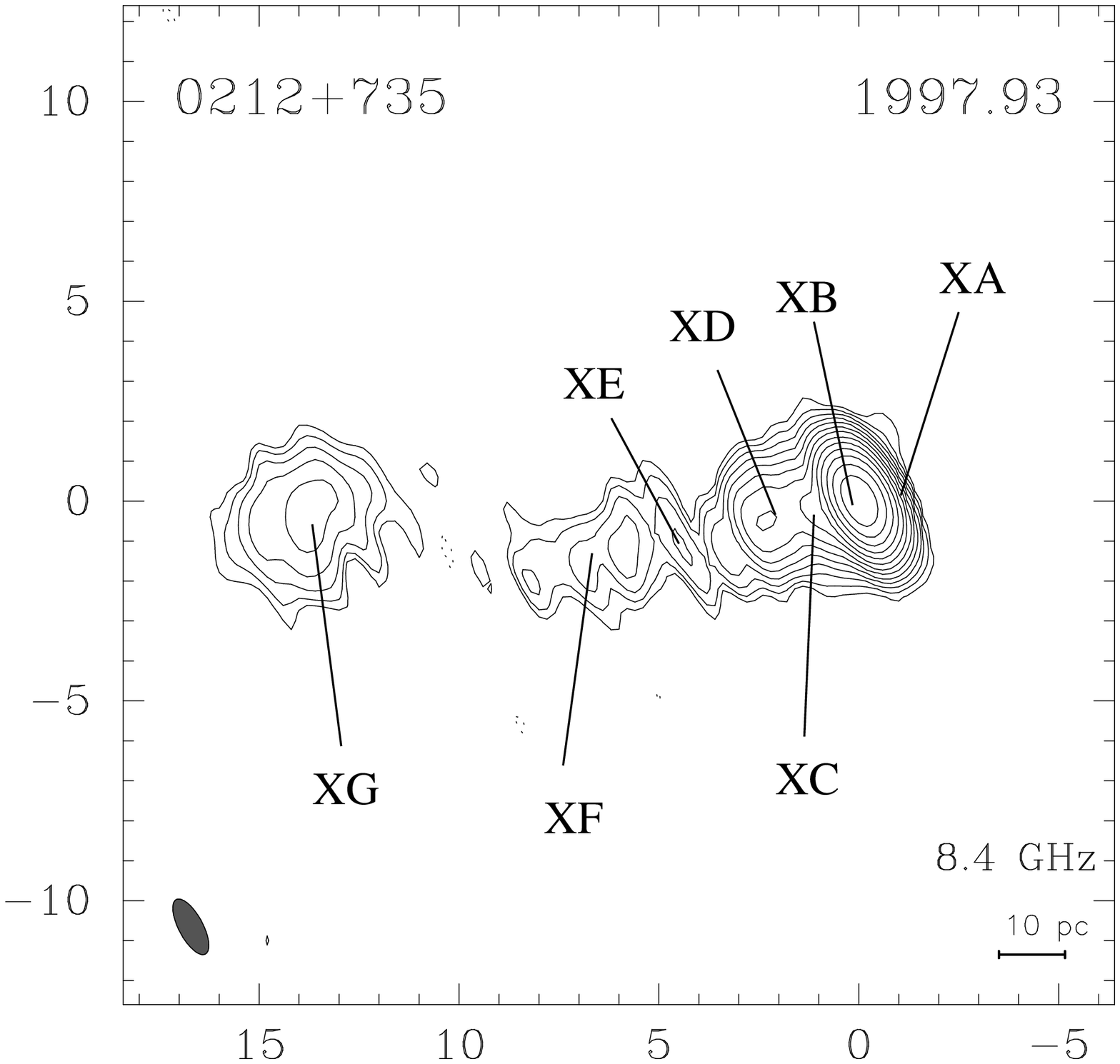}
\vspace{228pt}
\includegraphics{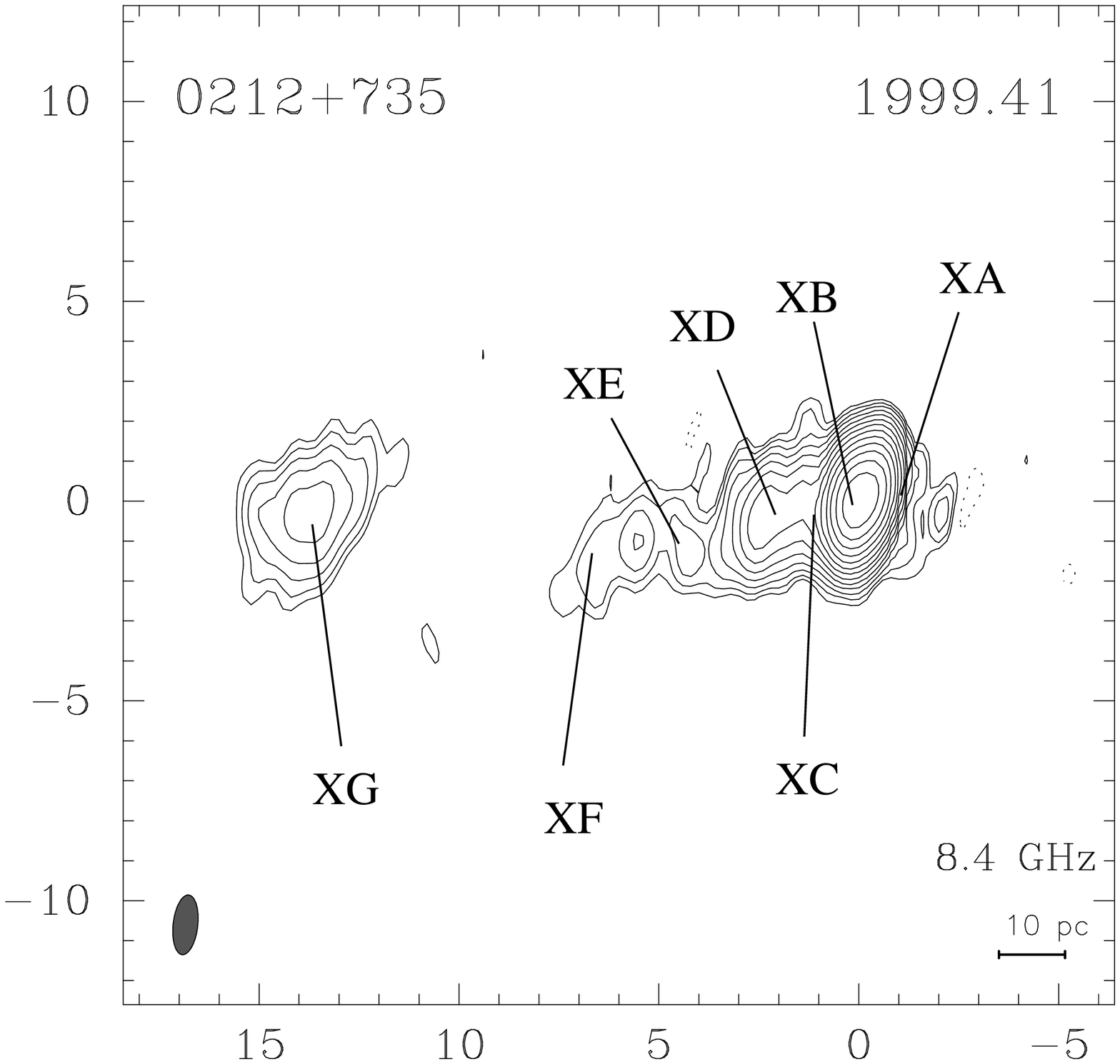}
\caption{VLBA images of \object{QSO\,0212+735}, observed on 6 December 1997
(1997.93) and 28 May 1999 (1999.41).  Axes are
relative $\alpha$ and $\delta$ in mas.  
See Table 1 for contour levels, 
synthesized beam size (bottom left in the figure), and
peak flux densities.
{See Table~\ref{table:modelfit} for component parametrization.}}
\label{fig:map0212}
\end{figure}

Our maps at 8.4\,GHz (Fig.~\ref{fig:map0212}) display a 
jet-like structure {prolonging} eastwards up to 14\,mas.
We reproduce the emission of the radio source at each epoch with a 7--component
model (Table~\ref{table:modelfit}). 
The peak of brightness corresponds to component XB. 
In the first epoch, XA is at $\sim$0.5\,mas {northwest}
of XB, in P.A.\,$\sim$55\degr.
In the second epoch, some
structure is visible west of the brightest region XA/XB.

%

Lacking an accurate astrometric registration, 
all components seem to have remained unchanged for the last 20 years.  
Note that component XG, about 13.8\,mas (P.A.\,$\sim$93\degr) east of 
the peak of
brightness, with a 3\% of the total mapped flux density, appears
unchanged between our two epochs.

\subsection{\object{BL\,0454+844}\label{subsec:0454}}

\begin{figure}[htbp]
\vspace{231pt}
\includegraphics{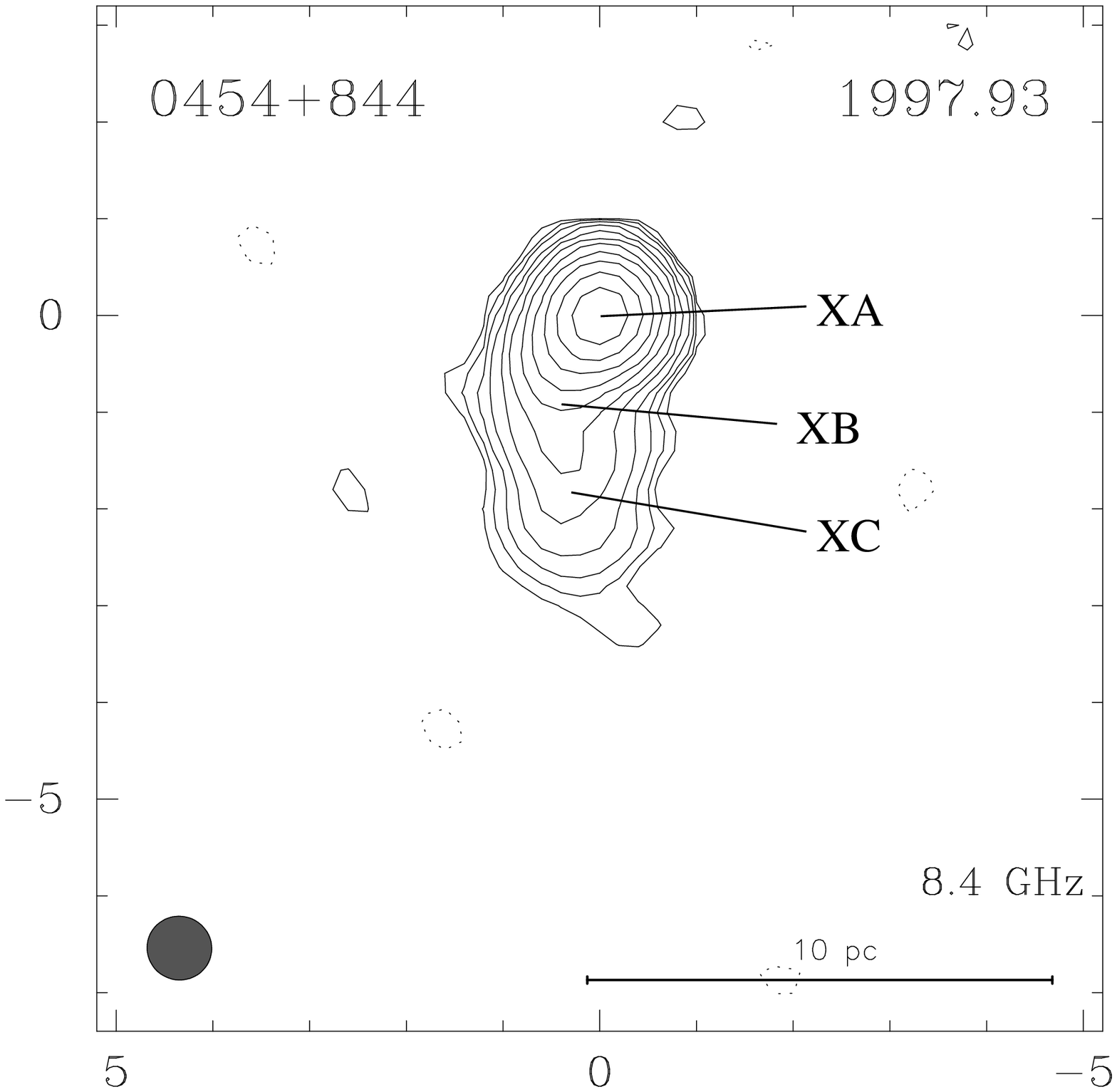}
\vspace{231pt}
\includegraphics{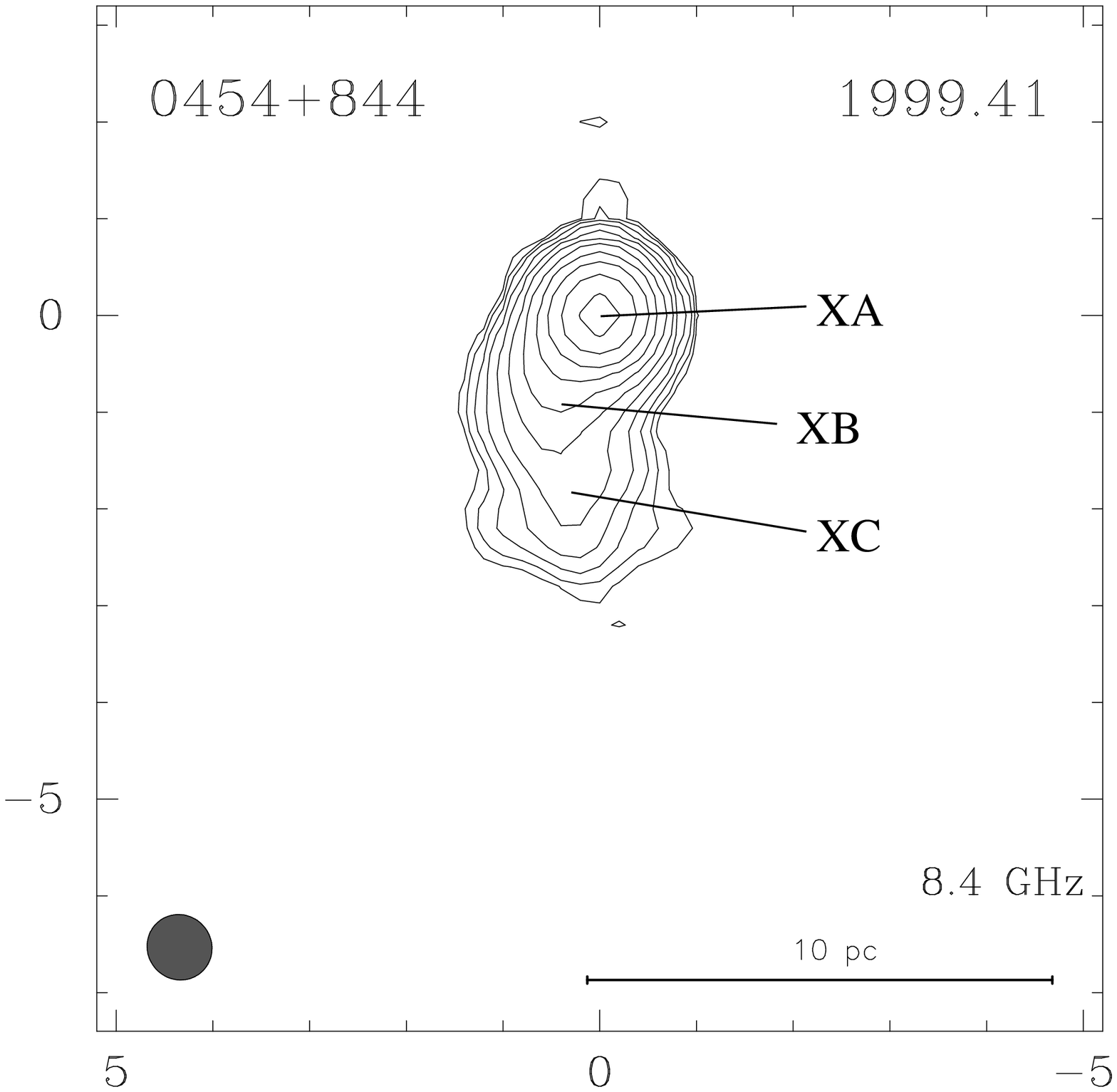}
\caption{VLBA images of \object{BL\,0454+844}, observed 
on 6 December 1997 (1997.93) and 28 May 1999 (1999.41).  Axes are
relative $\alpha$ and $\delta$ in mas. 
left.  
See Table 1 for contour levels, 
synthesized beam sizes (bottom left in the maps), and
peak flux densities.
{See Table~\ref{table:modelfit} for component parametrization.}}
\label{fig:map0454}
\end{figure}

The BL Lacertae object \object{BL\,0454+844}, with $V$=16.5 and $z$=0.112 (Stickel \& K\"uhr
\cite{sti96})
has been reported to contain all the flux density in its milliarcsecond
structure (Eckart \et\ \cite{eck87}).  Those authors reported a strong
core emission and weaker extended emission southwards, 
which was also confirmed by Pearson \& Readhead (\cite{pea88}).
Its radio emission shows variability on scales of (several) months.  
It registered a local maximum in mid 1995, and it has since then 
been decreasing (by $\sim$40\%) through late 1999. 
Then, its flux density has started to increase again 
(Peng \et\ \cite{pen00}).  
Our maps (Fig.\ \ref{fig:map0454})
show a very compact structure.  
There is no emission outside
$\sim$4\,mas of the core.  We reproduce its emission with a three-component
model (see Table \ref{table:modelfit}).
{It is interesting to notice
that the most compact source in the sample has also the smallest
redshift.}

%

\subsection{\object{QSO\,0615+820}\label{subsec:0615}}

The \object{QSO\,0615+820} is 
a radio source with $V$=17.5 and $z$=0.710 (Stickel 
\& K\"uhr \cite{sti96}). 
Eckart \et\ (\cite{eck87}) reported this radio source to be unresolved,
and modeled it with a single elliptical Gaussian profile elongated in 
{P.A.\,195\degr} at 5\,GHz and in {P.A.\,181\degr} at 15\,GHz. 

\begin{figure}[htbp]
\vspace*{228pt}
\includegraphics{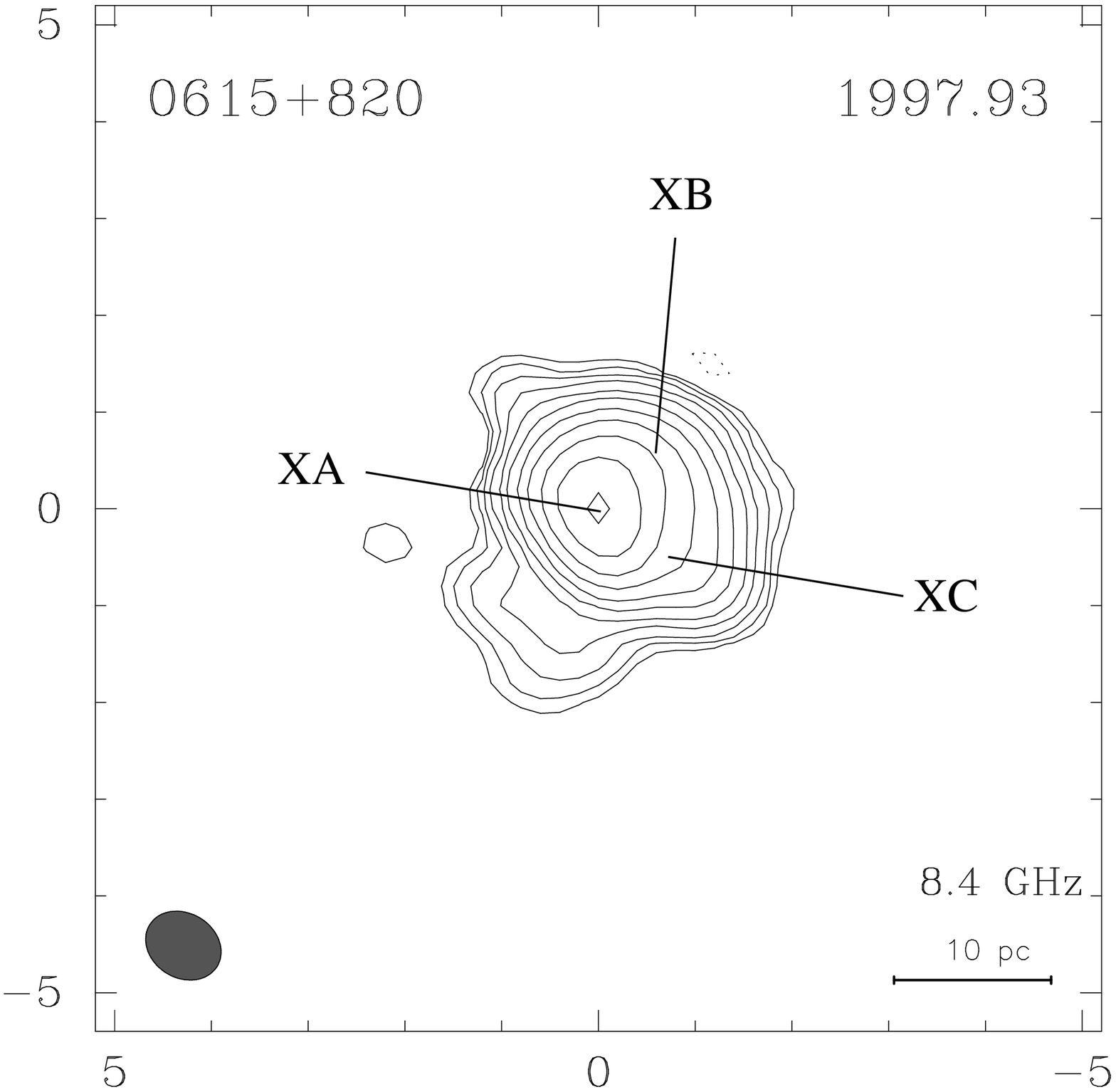}
\vspace*{228pt}
\includegraphics{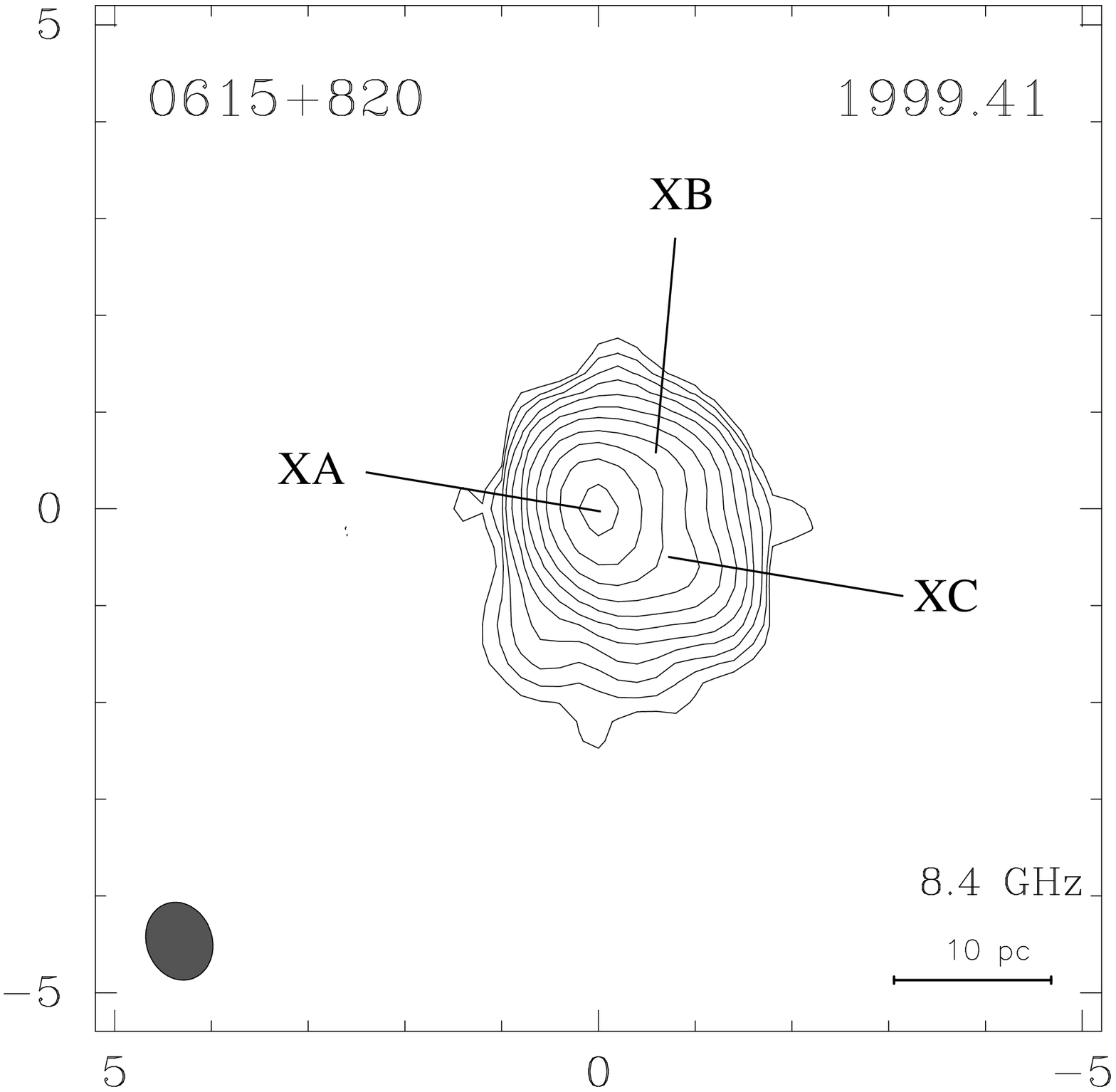}
\caption{VLBA images of \object{QSO\,0615+820}, observed 
on 6 December 1997 (1997.93) and 28 May 1999 (1999.41).  Axes are
relative $\alpha$ and $\delta$ in mas.  
See Table 1 for contour levels,
synthesized beam sizes (bottom left in the maps), and
peak flux densities.
{See Table~\ref{table:modelfit} for component parametrization.}}
\label{fig:map0615}
\end{figure}

Our maps (Fig.~\ref{fig:map0615})  show {a compact plateau-like 
structure} within 2--3\,mas;
however, we could not reasonably fit the brightness distribution
with a single elliptical Gaussian component. 
We obtained good Gaussian fits (Table~\ref{table:modelfit}) with three
close components within 1\,mas, with a spatial L-shaped distribution, 
being the brightest one XA, {to the east}.

%

\subsection{\object{BL\,0716+714}\label{subsec:0716}}

The object \object{BL\,0716+714} ($V=14.17$) 
is a rapidly variable source across the electromagnetic spectrum,
a paradigm of the intra-day variable (IDV) behavior
(Quirrenbach \et\ \cite{qui91}; Wagner \et\ \cite{wag96}).
No sign of optical counterpart exists, and thus no redshift is available.  
This fact {prompted} several authors (e.g.,\ Eckart \et\ \cite{eck87}) 
to assume a value of $z>0.3$.  Previous radio
maps show a core-halo structure on arcsecond scales (Antonucci \et\
\cite{ant86}).  VLBI maps at 5 and 22\,GHz show a very compact source on 
the milliarcsecond-scale with some hints of a core-jet structure oriented
along a P.A.\ $\sim$10\degr (Eckart \et\ \cite{eck87}).  

\begin{figure}[htbp]
\vspace*{228pt}
\includegraphics{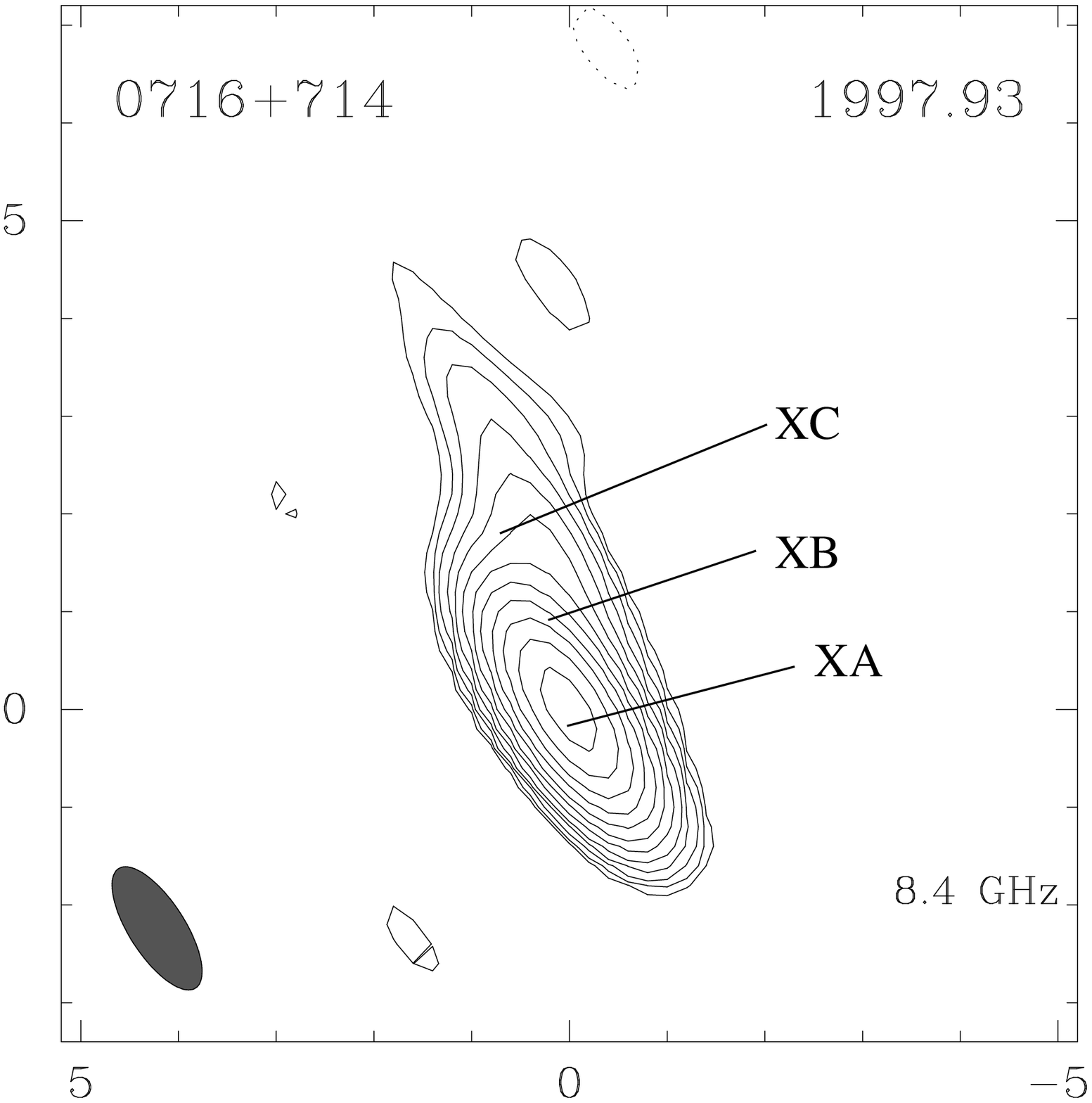}
\vspace*{228pt}
\includegraphics{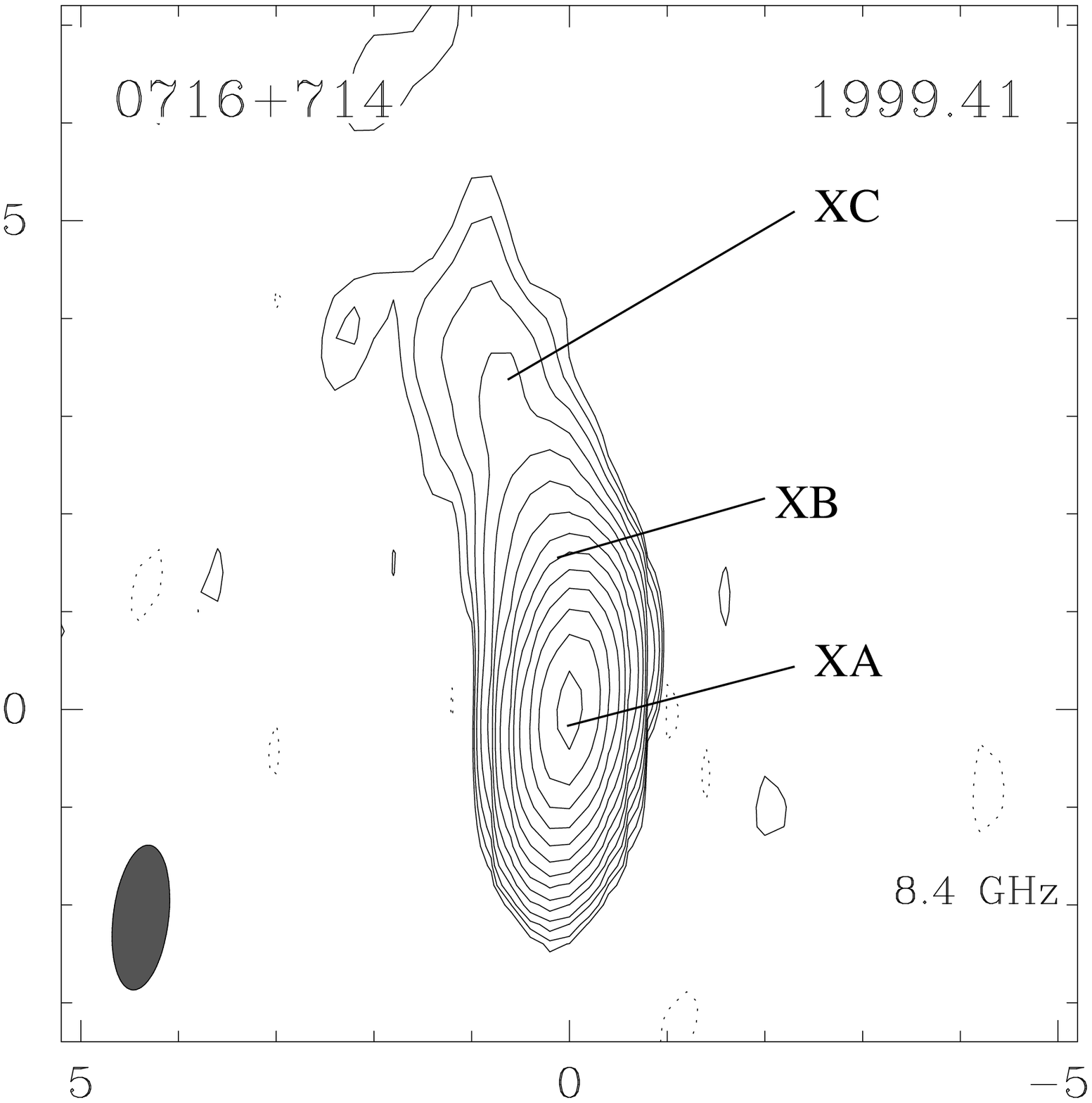}
\caption{VLBA images of \object{BL\,0716+714}, 
observed on 6 December 1997 (1997.93) and 28 May 1999 (1999.41).  Axes are
relative $\alpha$ and $\delta$ in mas.
See Table 1 for contour levels, synthesized beam sizes 
(bottom left in the maps), and
peak flux densities.
{See Table~\ref{table:modelfit} for component parametrization.}}
\label{fig:map0716}
\end{figure}

Our maps
(Fig.~\ref{fig:map0716}) show a core-jet structure extended northwards.
We modeled the structure
with three components (see Table \ref{table:modelfit}).  In the first
epoch the components lie at 0.8\,mas (XB) and 1.8\,mas (XC) 
(P.A.\,$\sim$11\degr) of the main component (XA),
and in the second epoch, at 1.0 and
3.3\,mas (again P.A.\,$\sim$11\degr).  
Taking the association at face value, and
for the elapsed period of time between both epochs ($\sim$1.5\,yr), 
XC would have moved with a rate of $\sim$0.7\,mas\,yr$^{-1}$. However,
\object{BL\,0716+714} looks more than a source 
with ``flaming" features than
the ``classical"
core-jet structure with components emerging from the core and
traveling along the jet.  In a {classical} core-jet source
the emergence of components is related to changes in the flux
density.  For this source, the
flux density has changed dramatically
from one epoch to the other (University of Michigan Radio Astronomy
Observatory data base and Peng \et\ \cite{pen00}).  
After smooth flux density changes from 1994 to 1997, the
flux density began to rise by mid 1997.  A decrease followed by
late 1997,
and a rise of more than
100\% by the first half of 1998, remaining stable 
{over the following} months
until a further decrease {took place} by mid 1999.
We mapped
a total flux density of 0.377\,Jy in 1997.93, and 0.990\,Jy in 1999.41.
This source is a case where frequent sampling of the structure and astrometric
registration are essential to correctly interpret the source structure
changes.

%

\subsection{\object{QSO\,0836+710} (\object{4C\,71.07})\label{subsec:0836}}

The \object{QSO\,0836+710} has $V$=16.5 and $z$=2.172 (Stickel \& K\"uhr
\cite{sti96}).  
At kpc-scales, it displays a highly polarized secondary component
{1.3\,arcsec off the core at} P.A.$\sim$200$^\circ$ (Perley \cite{per80},
O'Dea \et\ \cite{ode88}).  MERLIN-VLBI images
{at 90 and 18\,cm show a jet structure
extending up to 150\,mas in direction to the outer arcsecond lobe
(Hummel \et\ \cite{hum92b})}.  
It shows a complex 
and wiggled one-sided core-jet structure along 
P.A.$\sim$215$^\circ$ at {pc-scales.  The} structure of
the jet can be characterized by a sequence of kinks
among which the jet follows a slightly curved path
(Krichbaum \et\ \cite{kri90}).  
Based on data at 8.4, 15, 22, and 86 GHz, Otterbein \et\ (\cite{ott98}) 
reported the ejection of a new component at mas scales around epoch 1992.65,
with an apparent superluminal motion
of $\mu$=0.26$\pm$0.03\,mas/yr.
The ejection of this component would be directly related to {gamma-,}
X-ray, and optical activity observed in February 1992 (Otterbein
\et\ \cite{ott98}).

\begin{figure}[htbp]
\vspace*{225pt}
\includegraphics{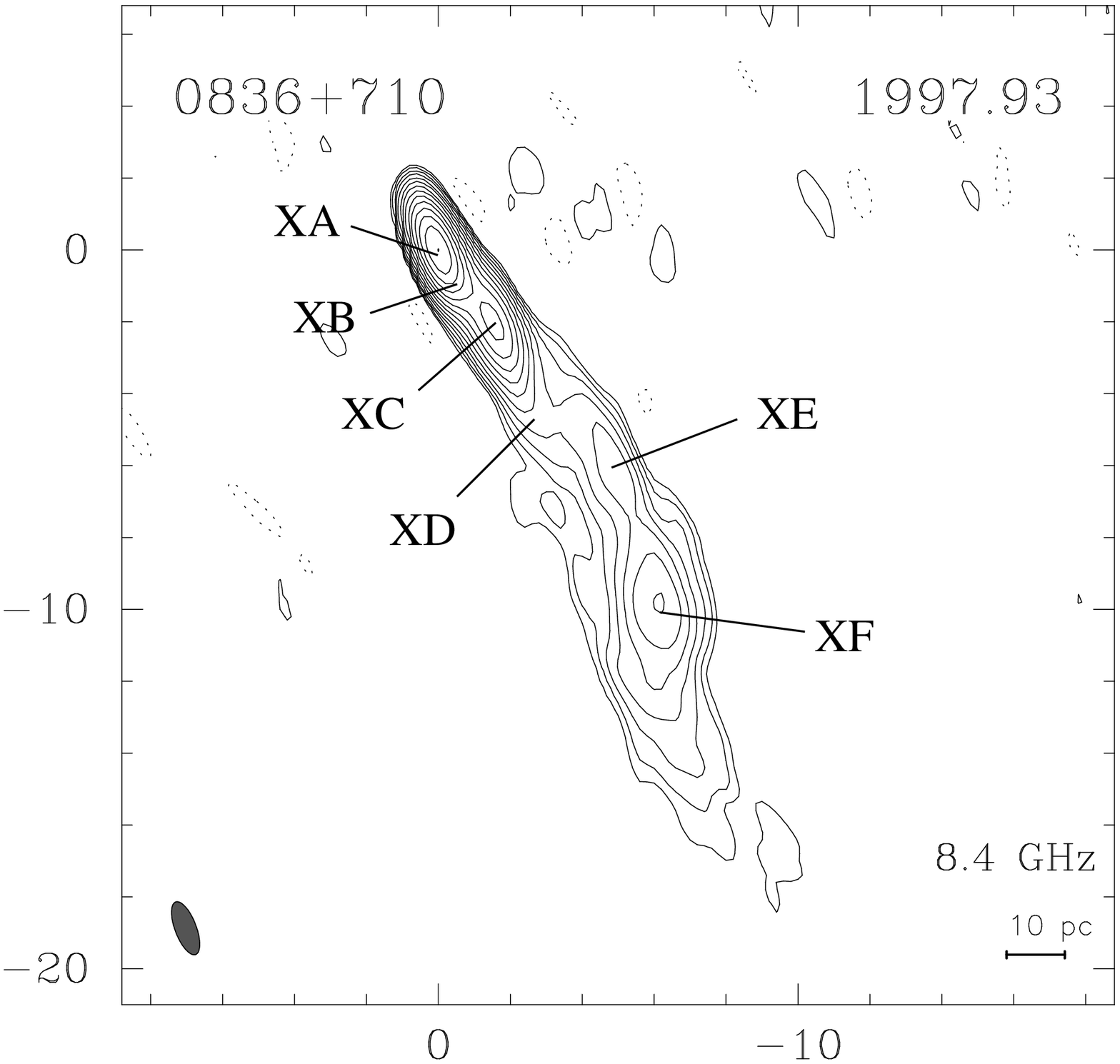}
\vspace*{225pt}
\includegraphics{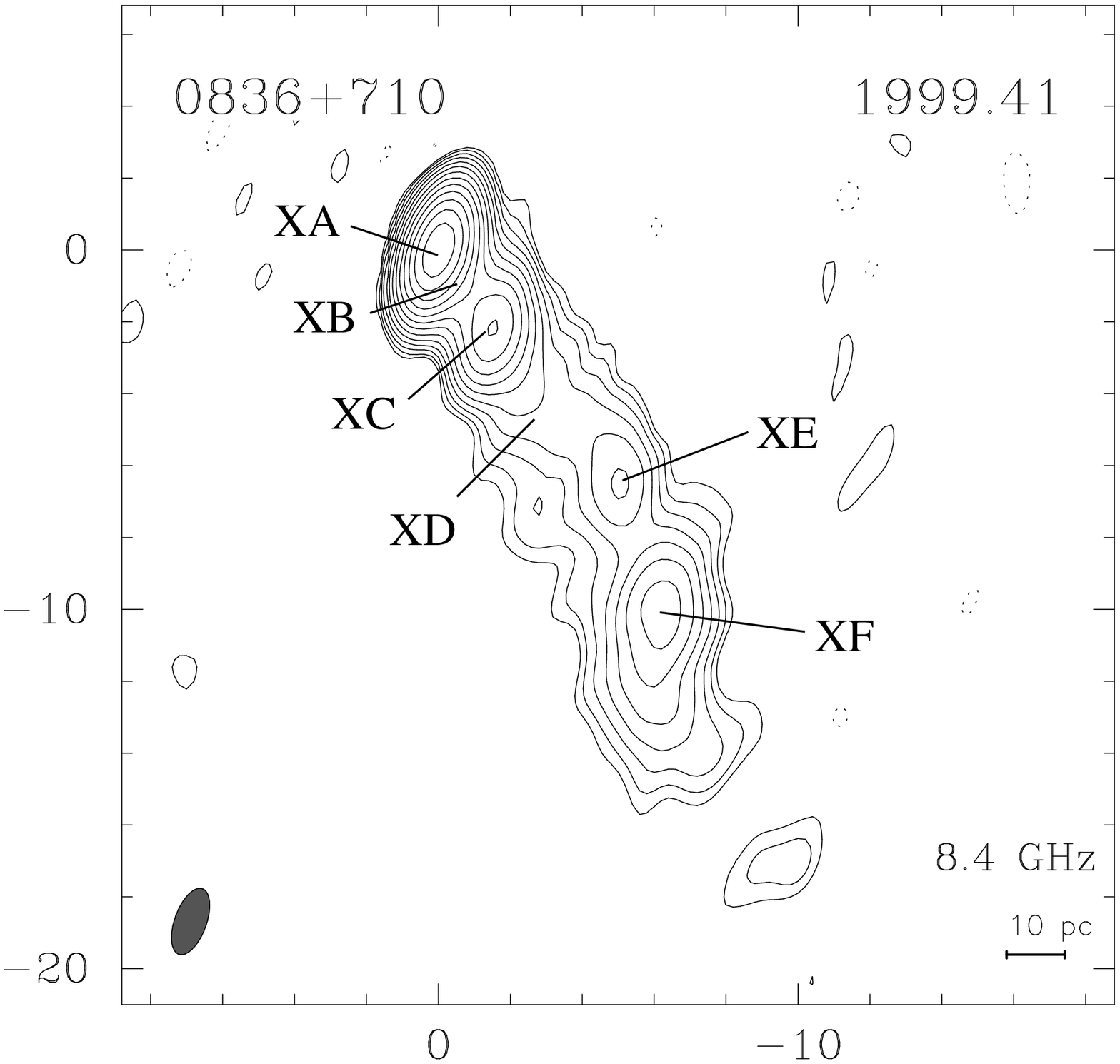}
\caption{VLBA images of \object{QSO\,0836+710}, 
observed on 6 December 1997 (1997.93) and 28 May 1999 (1999.41).  Axes are
relative $\alpha$ and $\delta$ in mas.
See Table 1 for contour levels,
synthesized beam sizes (bottom left in the maps), and
peak flux densities.
{See Table~\ref{table:modelfit} for component parametrization.}}
\label{fig:map0836}
\end{figure}

\begin{figure}[htbp]
\vspace{144pt}
\includegraphics{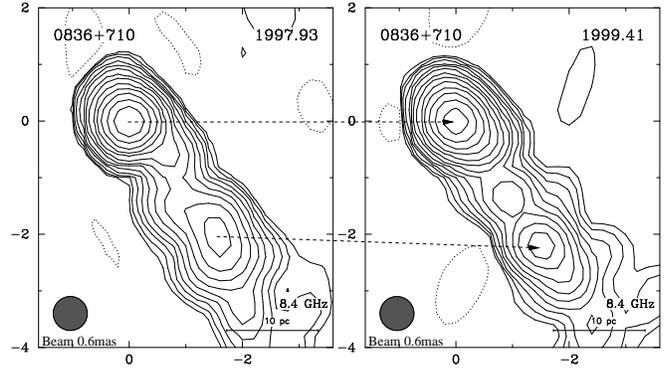}
\caption{VLBA images of \object{QSO\,0836+710},
convolved with a 0.6\,mas circular beam.  The map parameters are the
same as in Fig.\ \ref{fig:map0836}.  Note the small structural
changes in the inner part of the jet.  The peak of brightness changes from
1.233\,Jy/beam at 1997.93 to 1.050\,Jy/beam at 1999.41.  The dashed
lines draw a tentative association between features from one epoch to
another.
\label{fig:map0836-06}
}
\end{figure}

{Although} our maps 
(Fig.\ \ref{fig:map0836}) are convolved with differently 
oriented beams, the features in the maps can be identified unambiguously.
We detect emission up
to 30\,mas away of the core, although we plot the structure only 
{extending} 
15\,mas from the core to see more clearly the main features.
We reproduce the emission up to $\sim$12\,mas from the 
core with six components.
We convolved our images with a circular beam of 0.6\,mas size
to better compare the inner structure changes, and show
them in Fig.\ \ref{fig:map0836-06}.
We can neither confirm nor reject the superluminal motion reported by 
Otterbein \et\ (\cite{ott98}), until an astrometric registration 
between our both images becomes available. {Only from
our model fitting it
is difficult to establish proper motions between components  
from our model fitting (Table \ref{table:modelfit}).}

%

\subsection{\object{QSO\,1039+811}\label{subsec:1039}}

The \object{QSO\,1039+811} has $V$=16.5 and $z$=1.264 (Stickel \& K\"uhr
 \cite{sti96}).  Eckart \et\ (\cite{eck87}) reported
a pc-scale core-jet structure with a 
jet oriented with P.A.$\sim-70^\circ$
at 5\,GHz.  This radio source displays misalignment between the pc-scale and
the kpc-scale structure (Appl \et\ \cite{app96}).

\begin{figure}[htbp]
\vspace*{227pt}
\includegraphics{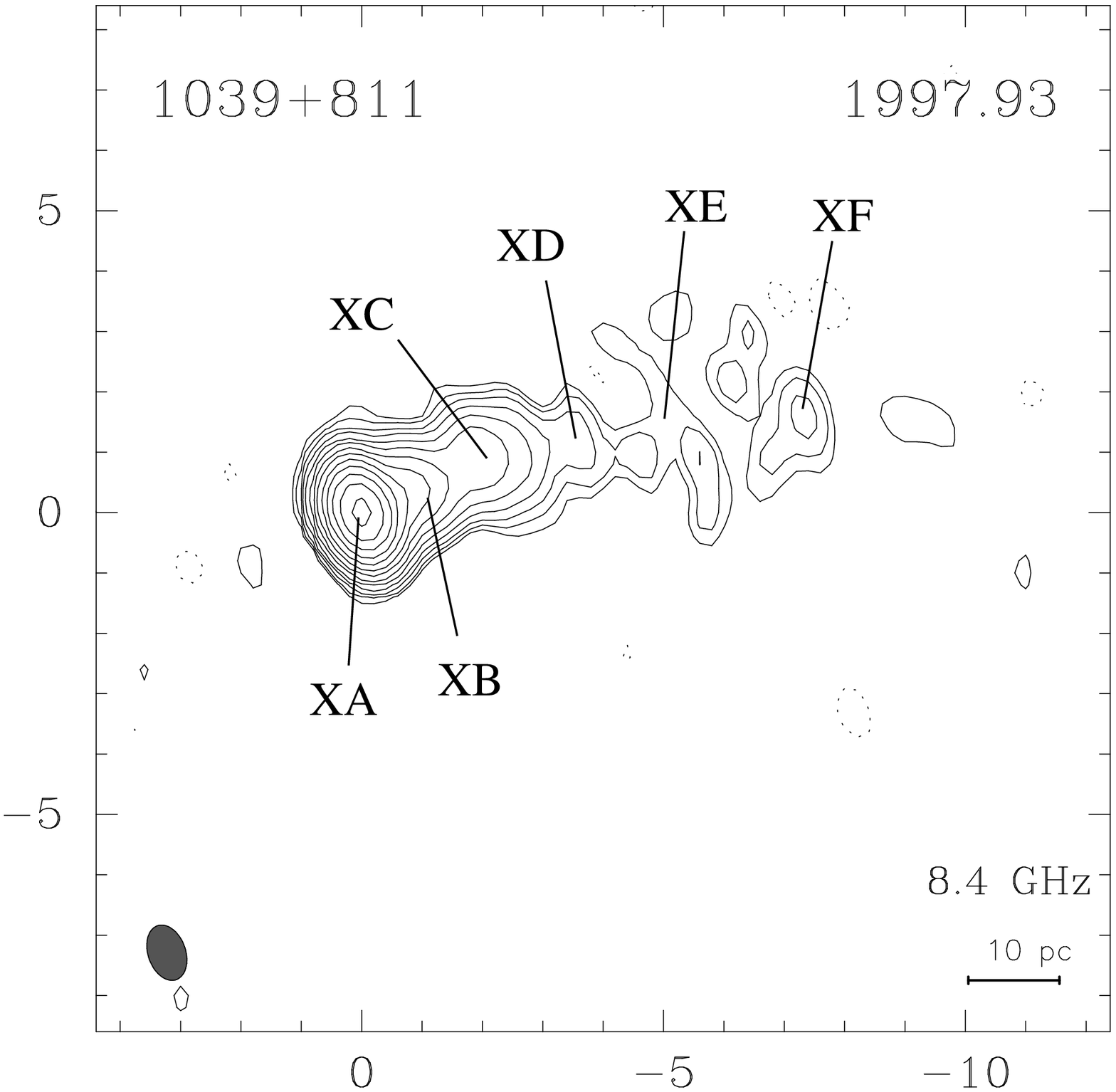}
\vspace*{227pt}
\includegraphics{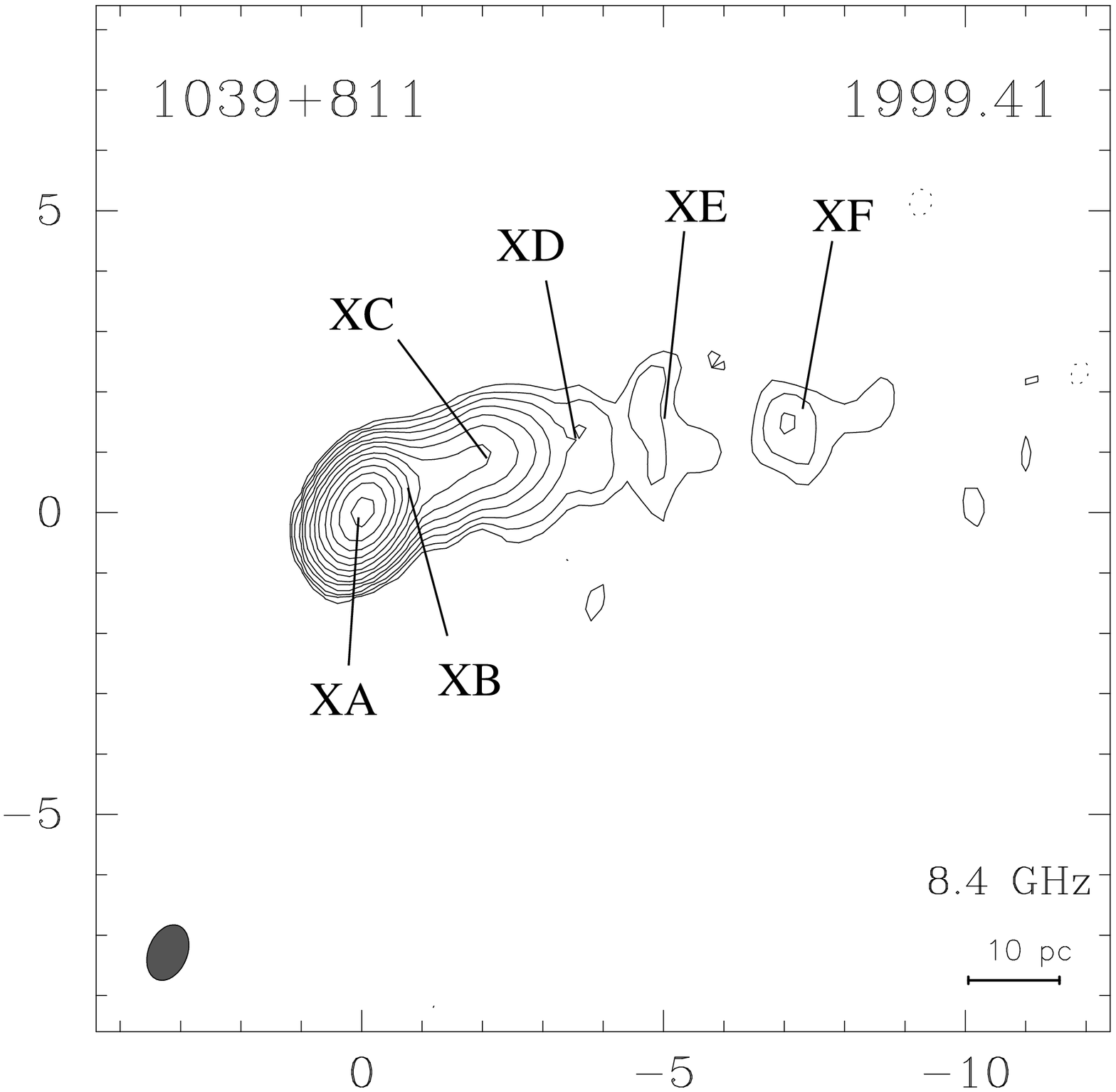}
\caption{VLBA images of \object{QSO\,1039+811}, 
observed on 6 December 1997 (1997.93) and 28 May 1999 (1999.41).  Axes are
relative $\alpha$ and $\delta$ in mas.
See Table 1 for contour levels,
synthesized beam sizes (bottom left in the figure), 
and peak flux densities.
{See Table~\ref{table:modelfit} for component parametrization.}}
\label{fig:map1039}
\end{figure}

Our maps (Fig.~\ref{fig:map1039})
show extended jet structure up to 10\,mas away of the {core} component.
The multi-frequency 
flux density monitoring reported by Peng \et\ (\cite{pen00}) shows a rise
from 1997 to 1999.  Our maps, however, have almost the same flux density
for both epochs (0.888 and 0.886\,Jy at 1997.93 and 1999.41,
respectively).  The model
fitting (Table \ref{table:modelfit}) reports an apparent 
backwards motion of the components XB (from 0.9 to 0.4\,mas from
XA), XC (2.1 to 1.8\,mas),
XD (3.1 to 2.5\,mas), XE (5.3 to 4.8\,mas), and XF (7.7 to 7.4\,mas).  
This apparently contracting motion
could be related to changes in the {core region}. The
emergence of a component, in its early {stage}, can produce apparent 
backward motions of the rest of the jet components at cm-wavelengths
(see, e.g., 
{Guirado \et\ (\cite{gui98}) and Ros \et\ (\cite{ros99}) for 
QSO\,1928+738}).
Those effects should be tested after the astrometric alignment of
the images becomes available.
Note that the moving components are very weak in comparison with the main
feature (which contains more than the 80\% of the mapped flux density in both
cases), and also that the $(u,v)$-sampling, and consequently the
synthesized beams, differ
{substantially} from the first epoch to the second.

%

\subsection{\object{QSO\,1150+812}\label{subsec:1150}}

The \object{QSO\,1150+812} has $V$=18.5 and $z$=1.250 (Stickel \& K\"uhr
 \cite{sti96}).  Its radio structure has been studied at pc and kpc scales,
by Appl \et\ (\cite{app96}), who reported a misalignment of  the jet
between both scales.
At kpc scales, Murphy \et\ (\cite{mur93}) reported a 
VLA map with a core and a fainter
component at 6 arcsec and P.A.$\sim -100$\degr.  
At pc scales, Eckart \et\ (\cite{eck87})
reported a core-jet structure in the VLBI maps, with a faint jet emission
oriented at P.A.$\sim$178$^\circ$, and superluminal {motion} of about 
0.12\,mas/yr of the jet component between 1979.93 and 1983.25 epochs.
Fey \& Charlot (\cite{fey97}) published {VLBI maps from 
epoch} 1995.20 
epoch at 8.4 and 2.3\,GHz.  {Extended} emission is 
detected up to 30\,mas (P.A.\ of 135$^\circ$) in their 2.3\,GHz maps. 
In their 8.4\,GHz maps, the {emission extends}
$\sim$7\,mas (P.A.$\sim$135$^\circ$).
P\'erez-Torres \et\ (\cite{per00})
studied this radio source astrometrically 
referenced to \object{BL\,1803+784}, and thus our 
observations constitute a second and third
astrometric epochs for the pair 
\object{QSO\,1150+812}/\object{BL\,1803+784}.  A comparison of the
astrometric results will be published elsewhere.

\begin{figure}[htbp]
\vspace*{227pt}
\includegraphics{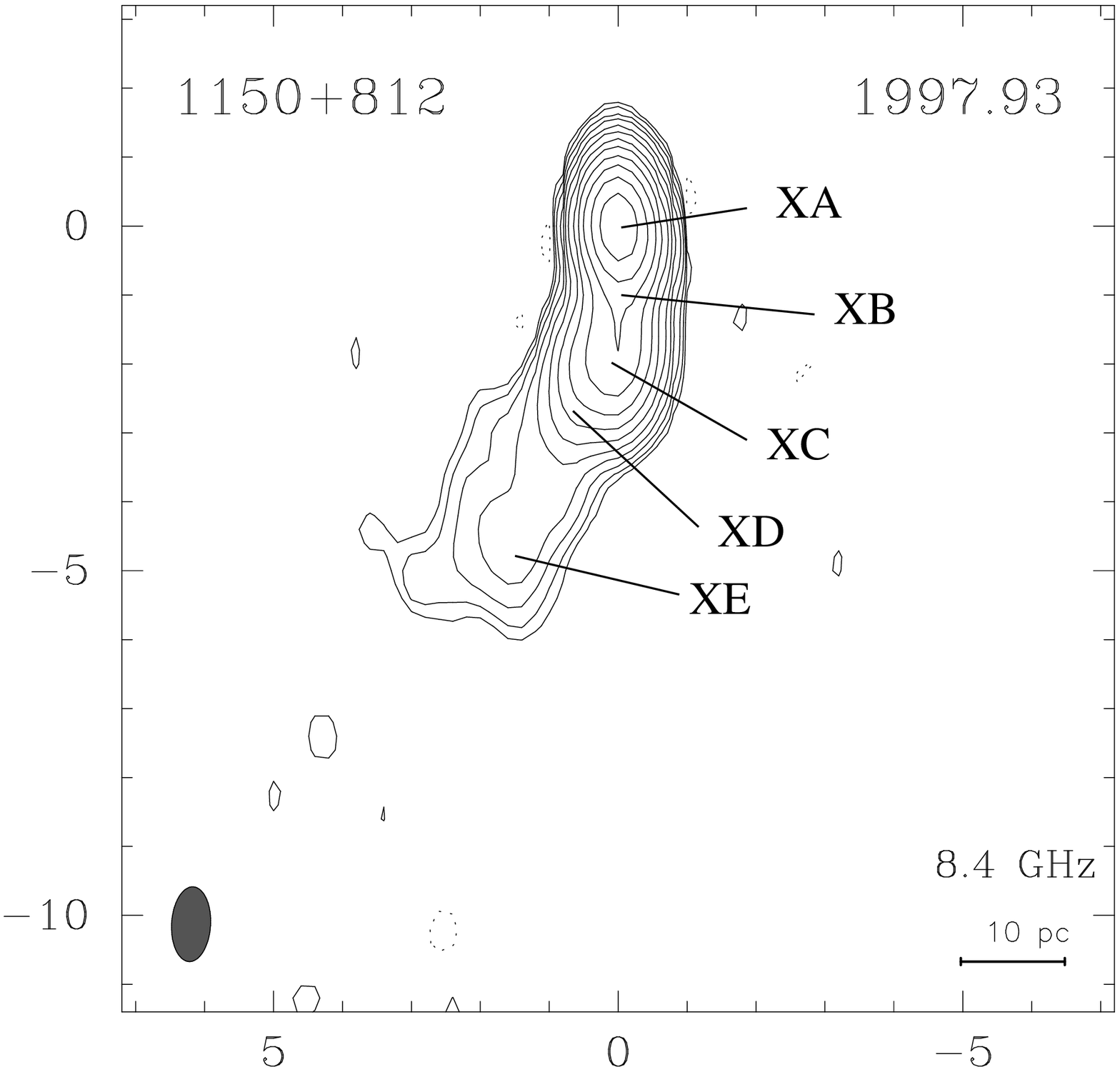}
\vspace*{227pt}
\includegraphics{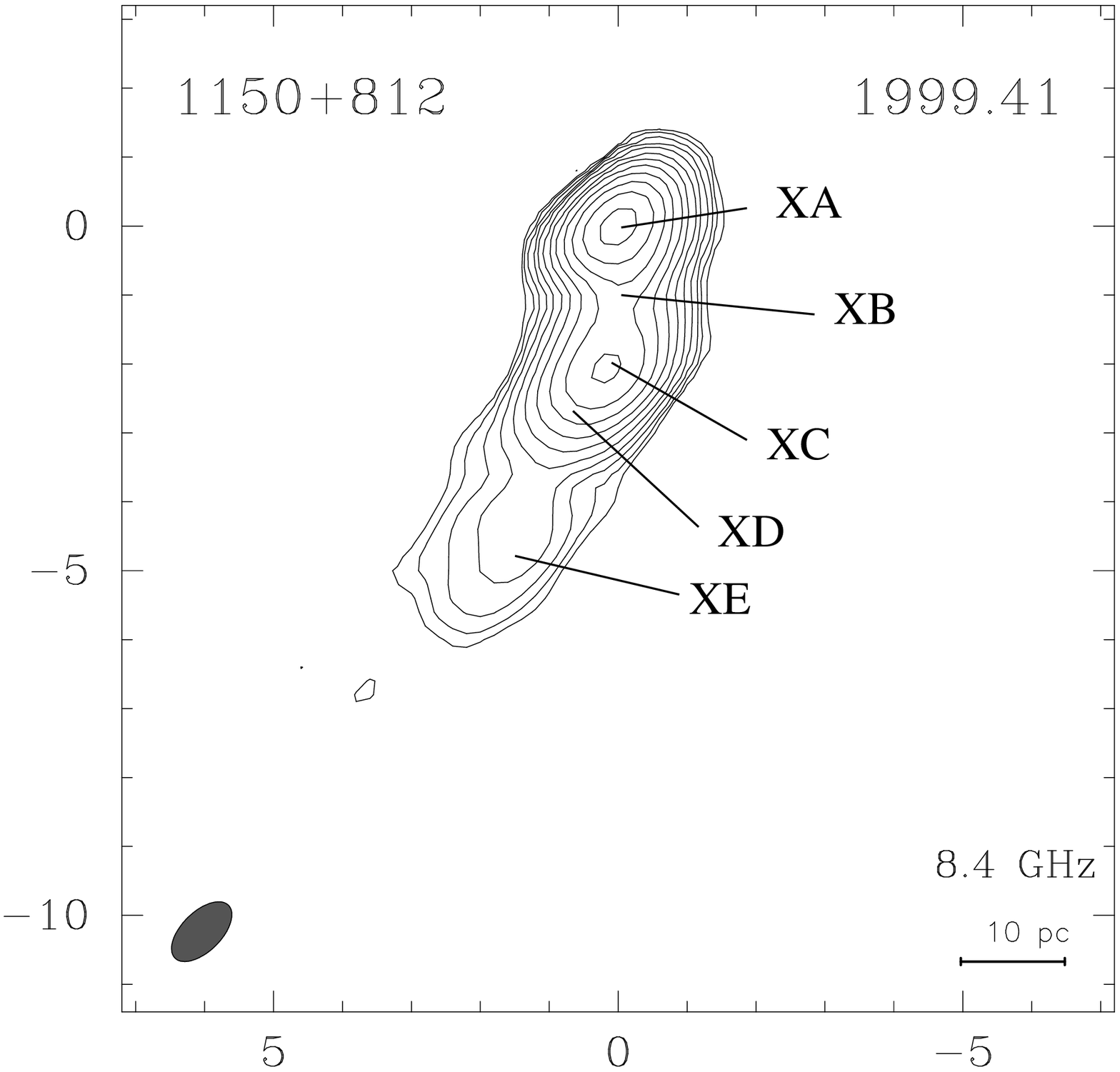}
\caption{VLBA images of \object{QSO\,1150+812}, observed on 
6 December 1997 (1997.93) and 28 May 1999 (1999.41).  Axes are
relative $\alpha$ and $\delta$ in mas.
See Table 1 for contour levels,
synthesized beam sizes (bottom left in the maps), and
peak flux densities.
{See Table~\ref{table:modelfit} for component parametrization.}}
\label{fig:map1150}
\end{figure}

Our maps (Fig.~\ref{fig:map1150})
display a jet oriented southwards, up to 5\,mas away from the {core}
component. We model fit the source (Table \ref{table:modelfit}),
with 5 components in both cases, up to 4.4\,mas south of
the radio brightest component XA
(P.A.\ $\sim$160\degr).  The source increased
its total flux density by less than a 20\% from one epoch to the other 
(Peng \et\ \cite{pen00}), and the VLBA flux {densities 
in our images are very similar}.
The radio source does not present important 
structural changes between {our} epochs, and the modelfitted components
do not permit to report any significant proper motion. 

%

\subsection{\object{BL\,1749+701}\label{subsec:1749}} 

The object \object{BL\,1749+701} has $V$=16.5 and $z$=0.770 (Stickel \& K\"uhr
 \cite{sti96}).  
VLA images at 5\,GHz (Perley \cite{per82}) show a halo of $\sim$0.4 arcsec.  
O'Dea \et\ (\cite{ode88}) reported two components at pc-scales at 
15\,GHz separated
by {0$\rlap{.}^{\prime \prime}$4} at a P.A.\,$-150^\circ$.
Kollgaard \et\ (\cite{kol92}) did not report extended emission, 
although they found evidence of emission to the north.  
VLBI observations from Eckart \et\ (\cite{eck87})
show the core, and a component at $\sim$2.4\,mas and {P.A.\,--45\degr}.
Witzel \et\ (\cite{wit88}) reported this component to be moving with
a rate of 0.1\,mas/yr.  B{\aa}{\aa}th \& Zhang (\cite{baa92}) reported
a more complicated jet structure, with position angle of $\sim$$-65^\circ$,
and gave a three-component model for maps from 1980 to 1983.  They
found a maximum component motion of 0.18\,mas/yr.  
Gabuzda \et\ (\cite{gab92}) reported
new results from epoch 1987, and give a 4-component model, at separations
of 0.84, 1.93, and 3.41\,mas of the core component (P.A.\ of $-60.7^\circ$,
$-66.5^\circ$, and $-53.0^\circ$, respectively).  
These authors claimed that the
radio source has an unusually low degree of polarization
for BL Lacertae objects,  and also questioned 
some of the proper motions reported above.

\begin{figure}[htbp]
\vspace*{225pt}
\includegraphics{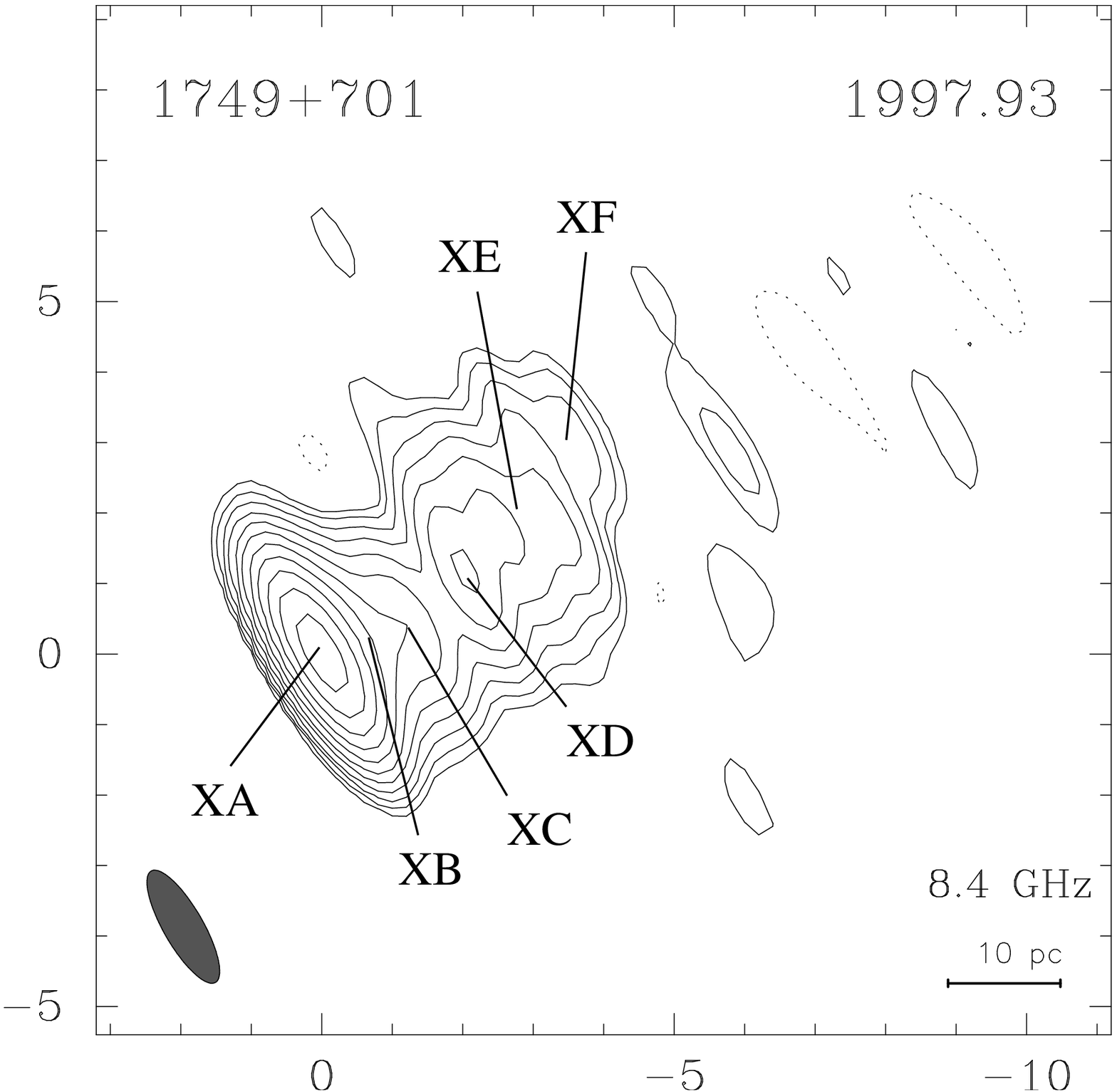}
\vspace*{225pt}
\includegraphics{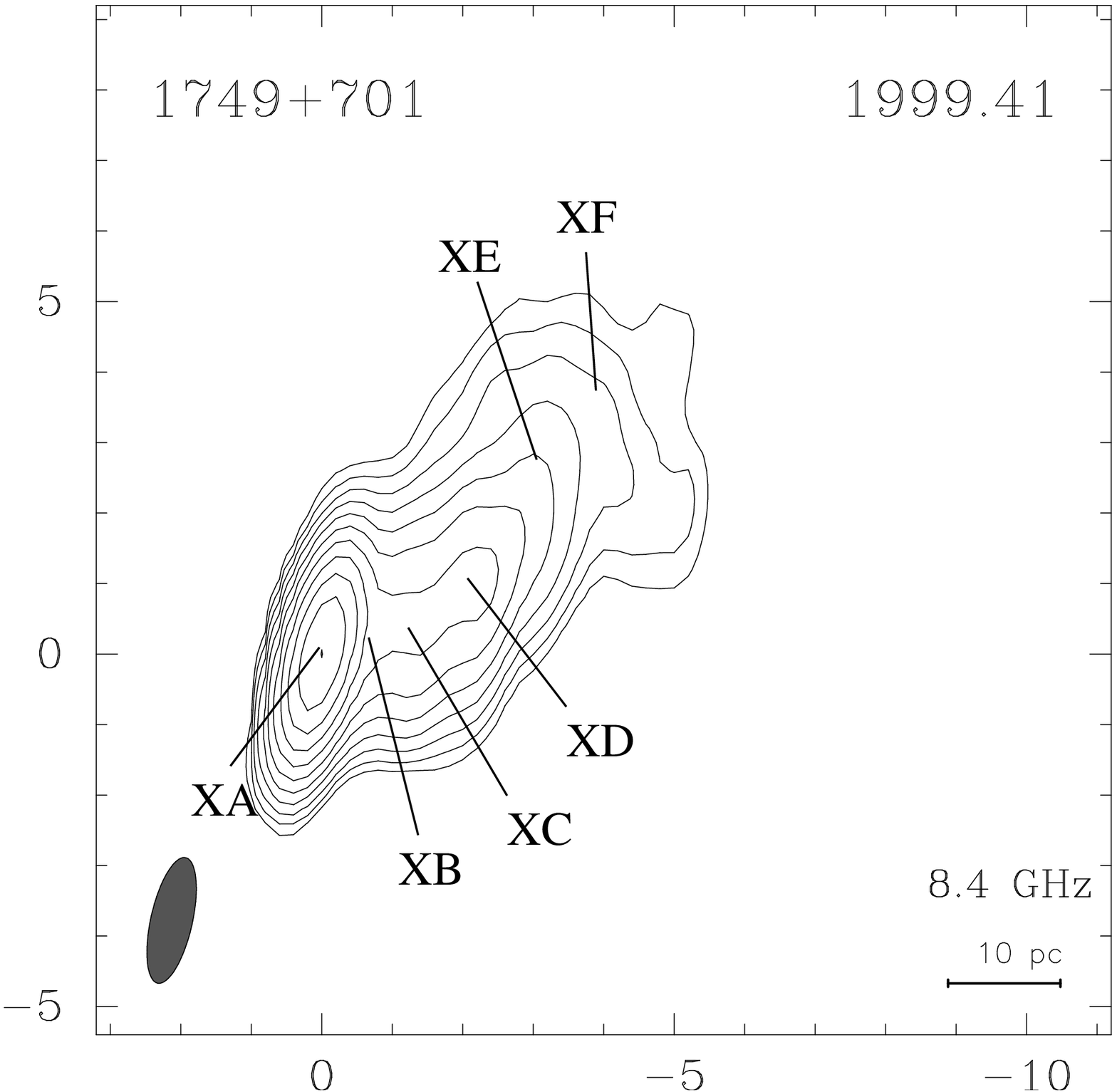}
\caption{VLBA images of \object{BL\,1749+701}, 
observed on 6 December 1997 (1997.93) and 28 May 1999 (1999.41).  Axes are
relative $\alpha$ and $\delta$ in mas.
See Table 1 for contour levels, 
synthesized beam size (bottom left in the maps), and
peak flux densities.
{See Table~\ref{table:modelfit} for component parametrization.}}
\label{fig:map1749}
\end{figure}

Our maps (Fig.~\ref{fig:map1749})
show a complex jet structure oriented to the northwest,
and extended up to 7\,mas from the core.
We identify in our images {more components than the 
reported by (B{\aa}{\aa}th \& Zhang \cite{baa92})}.
{We observe about the same structure for the two epochs ever being
the beam perpendicular to the jet direction
in the first epoch, and 
(unfortunately) more aligned to it
for the second one.}
A Gaussian model of the visibilities (see Table~\ref{table:modelfit})
includes a compact, central component (XA), and 5 components, XB to
XF, stretching {up $\sim$5\,mas 
from} XA, and describing a slightly
curved trajectory northwards.
This source offers interesting prospects of study
since not only the proper motions reported earlier will be unambiguously
checked, but also the astrometry will provide insight of the nature of 
the XA component, and the possible effects of opacity.

%

\subsection{\object{BL\,1803+784}\label{subsec:1803}}

The object \object{BL\,1803+784} has $V$=16.4 and $z$=0.864
(Stickel \& K\"uhr
\cite{sti96}).
It has been extensively studied in the past at different 
radio wavelengths.
At kpc scales, the maps reveal components south of the core 
at 2, 37, and 45\,arcsec (Antonucci \et\ \cite{ant86}, 
Strom \& Biermann \cite{str91}, 
Murphy \et\ \cite{mur93}).  Eckart \et\ (\cite{eck87}) and
Pearson \& Readhead (\cite{pea88}) reported a core-jet
structure at pc scales with a jet oriented westwards.  
Schalinski 
(\cite{sch90}) suggests the stationarity of a component at 1.2\,mas of
the core.  Krichbaum \et\ (\cite{kri93}, \cite{kri94}) with
{43\,GHz observations} reported the existence of some
traveling components between the core and {such component, now
estimated to be at}
1.4\,mas separation.
Steffen (\cite{ste94}) modeled the radio source as an homogeneous
plasma jet of helical trajectory in adiabatic expansion.
Fey \et\ (\cite{fey96}) reported VLBA results at 8.5 and
2.3\,GHz.  At 8.5 GHz, the radio source displays {a 
12\,mas jet}.  At 2.3\,GHz, the radio source shows a jet up to 
40\,mas with P.A.$\sim-110^\circ$.  At this distance, the jet turns
northwards, reaching the extended emission a P.A.\ up 
to $-60^\circ$.
This radio source has been studied
astrometrically by Ros et al.\ (\cite{ros99}) with reference to 
\object{QSO\,1928+738}/\object{BL\,2007+777}, 
and by P\'erez-Torres et al.\ (\cite{per00}) with 
reference to \object{QSO\,1150+812}.  

\begin{figure}[htbp]
\vspace*{225pt}
\includegraphics{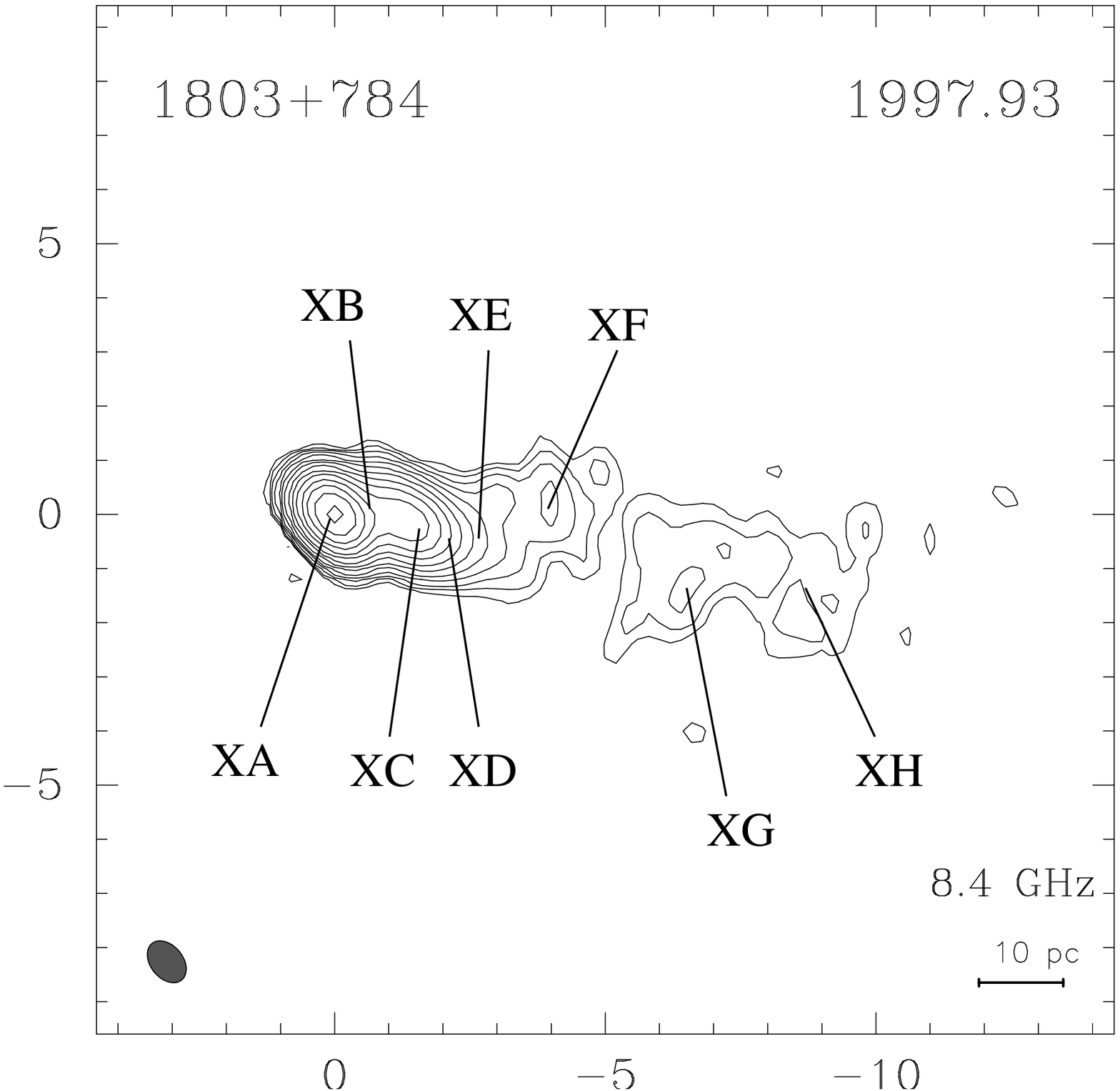}
\vspace*{225pt}
\includegraphics{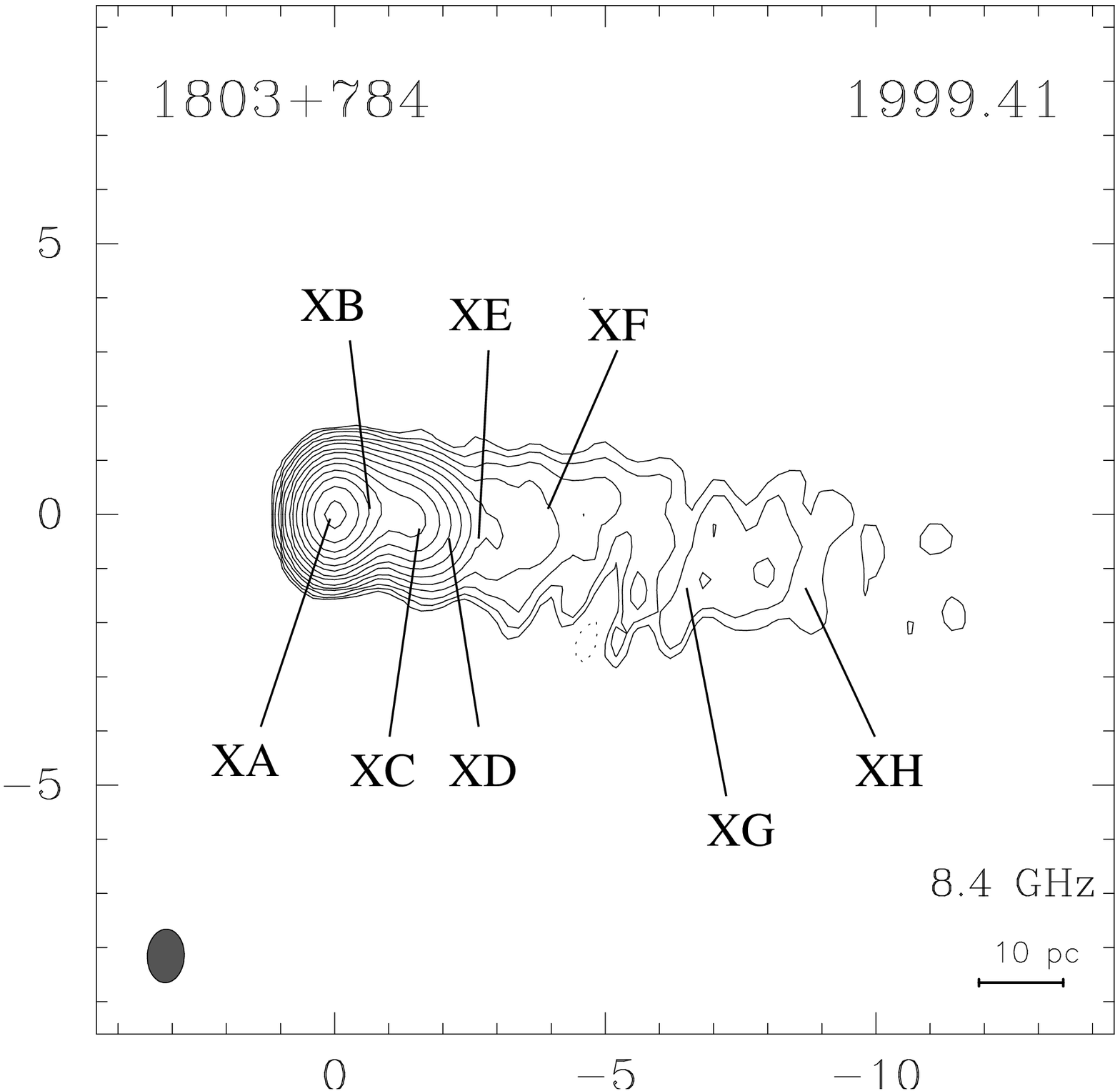}
\caption{VLBA images of \object{BL\,1803+784}, 
observed on 6 December 1997 (1997.93) and 28 May 1999 (1999.41).  Axes are
relative $\alpha$ and $\delta$ in mas.
Contours are 0.859\,mJy/beam $\times$ (1, 2, 4, ..., 1024).
See Table 1 for contour levels, synthesized beam sizes 
(bottom left in the maps), and
peak flux densities.
{See Table~\ref{table:modelfit} for component parametrization.}}
\label{fig:map1803}
\end{figure}

Our maps (Fig.~\ref{fig:map1803})  show similar features to those obtained
by Fey \et\ (\cite{fey96}).  The component at 1.2--1.4\,mas of the strong
core is present, as it has been for all VLBI observations during 
the last 20 years.  A model fitting of the visibilities with
8 elliptical Gaussian components (Table \ref{table:modelfit})
reproduces the structure of the radio source.
The inner region has three main components:
XA (with 1.5\,Jy, the ``core"), XB (at 
0.5--0.6\,mas with P.A.\ of $\sim-85$\degr), corresponding to the component
reported by Krichbaum \et\ \cite{kri93}), and XC (1.2\,mas, P.A.\ 
$\sim-95$\degr). Those three components represent $\sim$95\% of the
flux density of the source.
Further components range from $\sim$2 to 
$\sim$9\,mas from XA.  The jet bends to the south at $\sim$5\,mas of the
main feature.  At lower frequencies, the source has more extended
emission at  P.A.\,$\sim$$-100$\degr.  Such extended
emission is also present in our maps,
though not shown in Fig.~\ref{fig:map1803}. 
We find no evidence for {\em bona fide} proper motions of
the components from our two observing epochs.

%

\subsection{\object{QSO\,1928+738} (\object{4C\,73.18})\label{subsec:1928}}

\object{QSO\,1928+738} 
has $V$=15.5 and $z$=0.3021 (Roos \et\ \cite{roo93}), 
and displays {superluminal motion in a jet oriented westwards}. 
It is the best known and most extensively studied radio source of
the complete S5 polar cap sample.  At kpc scales, it displays extended emission
in {the north-south} direction, with two lobes separated 40\,arcsec 
(P.A.$\sim$0\degr) (Rusk \& Rusk \cite{rus86}).  
Johnston \et\ (\cite{joh87})  
and
Murphy \et\ (\cite{mur93})
detected emission up to 80\,arcsec from the core.  
Based on kpc-scale images, Hummel \et\ (\cite{hum92a}) 
modeled the radio source as an object with
a helical magnetic field attached to a rotating accretion disk.  
Eckart \et\
(\cite{eck85}) reported a self-absorbed core and a pc-scale jet 17\,mas 
long, whose components display superluminal motion.  
Roos \et\ ({\cite{roo93})
modeled the radio source as a binary black hole.  
Guirado \et\ (\cite{gui95a,gui98}) detected a shift in the peak of
brightness between different epochs, based on an astrometric analysis. 
Ros \et\ (\cite{ros99}) confirmed the results of 
Guirado et al., and  also studied the proper motions of 
\object{QSO\,1928+738} 
by aligning {astrometrically 
maps} of 1985 and 1991. 

Our maps (Fig.\ \ref{fig:map1928}) show very elliptical synthesized beams,
which nevertheless allow to distinguish perfectly different features
of its pc-scale. The flux density of
\object{QSO\,1928+738} 
at 8.4\,GHz has not changed much 
from the first to the second epoch.
{Hummel \et\ \cite{hum92a} reported that 
the core of this radio source ejects a new 
component every $\sim$1.6\,yr (approximately the time elapsed 
between our two observing epochs).}
If we convolve the {\sc clean}ed components with a 0.6\,mas circular beam
(Fig.\ \ref{fig:map1928-06}), structural changes are evident between 
the two epochs.
The component XA (north of the brightest one in 1997.93), 
has $\sim$0.8\,Jy, and XB (the brightest in 1997.93), 
has $\sim$1.3\,Jy.  In 1999.41 they have $\sim$0.9 and 
$\sim$0.6\,Jy, respectively.  Previous astrometric results (Guirado 
\et\ \cite{gui98}, Ros \et\ \cite{ros99}) show that most likely none of
those components {corresponds to the true ``core''}, which should be 
northwards of them (shown as a question mark 
in Fig.\ \ref{fig:map1928-06}) {and whose radiation is probably
very self-absorbed}.
If we take as reference component XB (Table \ref{table:modelfit}),
the rest of the components extend southwards in P.A.\,165\degr\, 
to 4--8\,mas distance and in 175\degr\, to 15--20\,mas.  
Though we could try to identify the components seen in
1991.88 by Ros \et\ (\cite{ros99})
with those reported for our first epoch ($\sim$7.1 years later),
such an attempt would be, at least, adventurous.
Indeed, for an emergence rate of one component every 1.6 years, 
\object{QSO\,1928+738} should have emerged 4 or 5 components,
making completely ambiguous any component identification.
The comparison of these 
components with results from previous epochs by us and by other
authors will be presented elsewhere.

%
\begin{figure}[htbp]
\vspace*{222pt}
\includegraphics{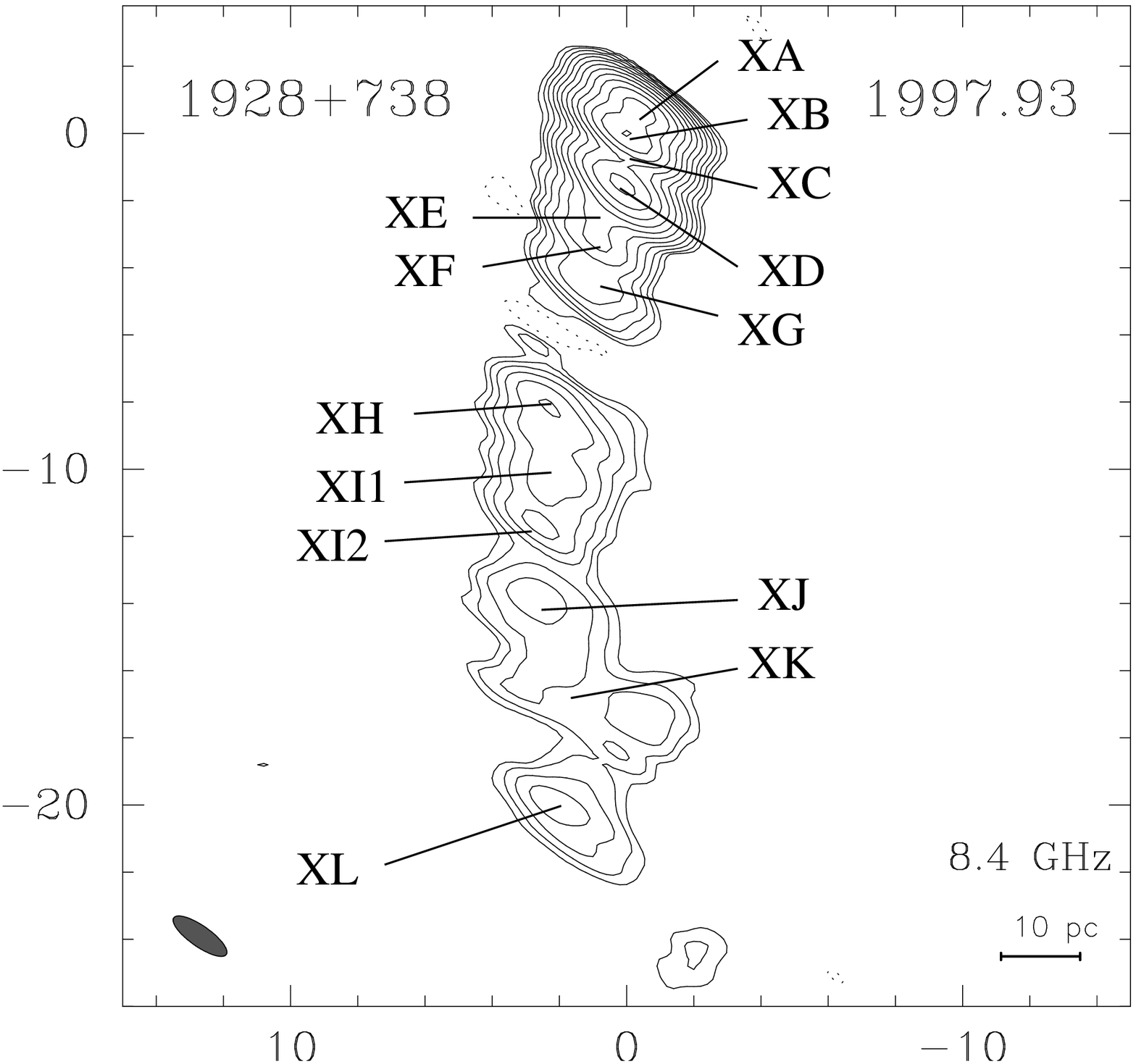}
\vspace*{222pt}
\includegraphics{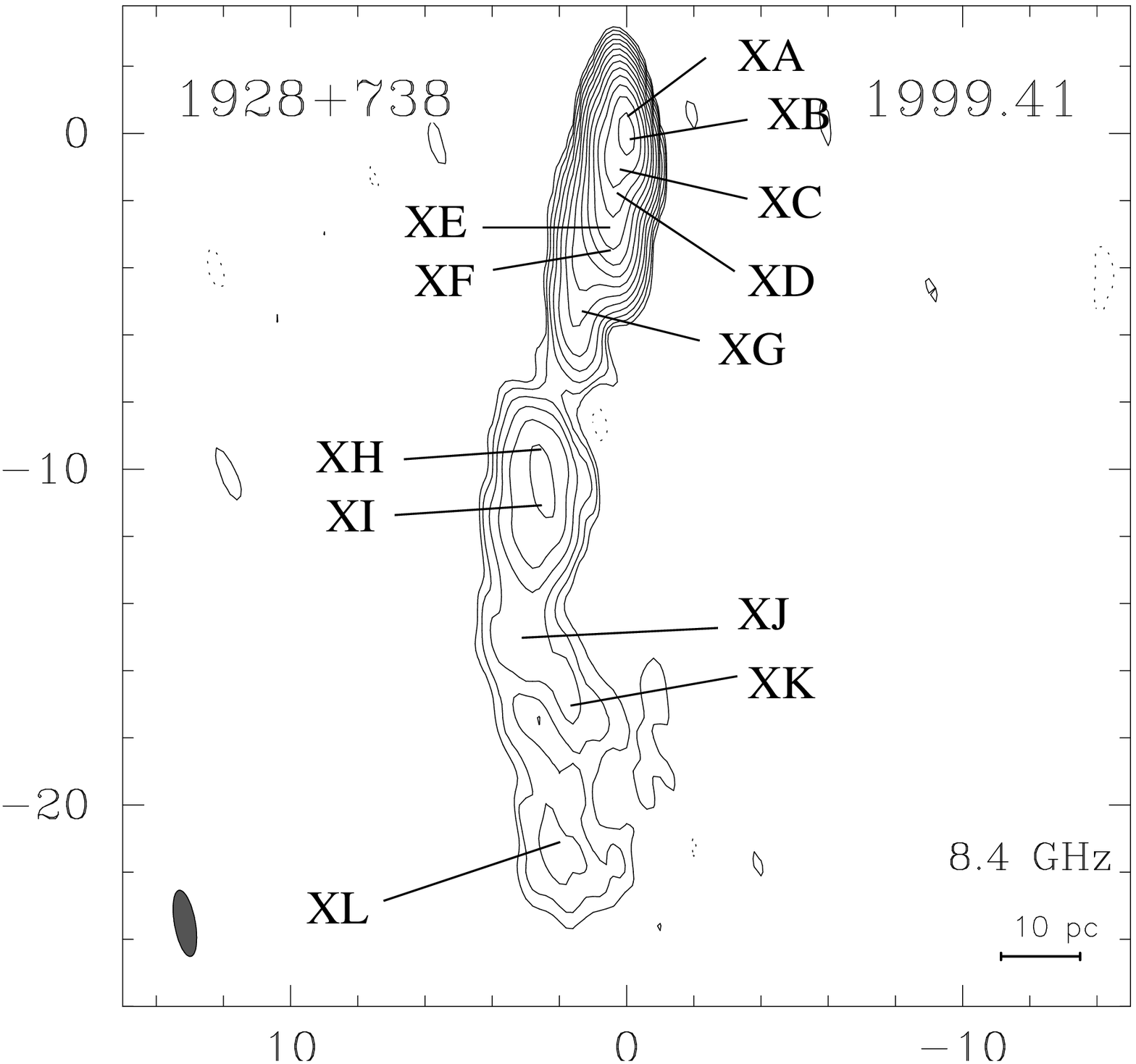}
\caption{VLBA images of \object{QSO\,1928+738} (\object{4C\,73.18}), 
observed on 6 December 1997 (1997.93) and 28 May 1999 (1999.41).  
Axes are relative $\alpha$ and $\delta$ in mas.
See Table 1 for contour levels,
synthesized beam sizes (bottom left in the maps), 
and peak flux densities.
{See Table~\ref{table:modelfit} for component parametrization.}}
\label{fig:map1928}
\end{figure}

\begin{figure}[htbp]
\vspace{159pt}
\includegraphics{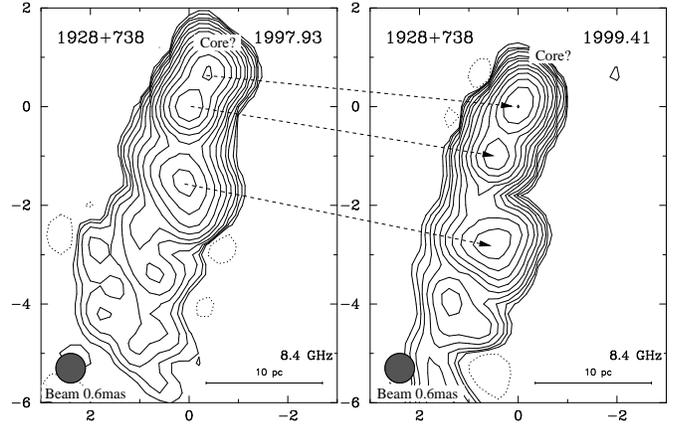}
\caption{VLBA images of \object{QSO\,1928+738}, convolved with a 0.6\,mas
circular beam.  Contours are the same as in 
Fig.\ \ref{fig:map1928}.  The dashed lines represent a tentative association
between jet regions from one epoch to another.  
Notice the structural changes in the maps, especially in the 
brighter regions.  The images are
centered in the brightest feature in both cases, which very probably diverges
from a rigorous astrometric registration.
\label{fig:map1928-06}
}
\end{figure}

\subsection{\object{BL\,2007+777}\label{subsec:2007}}

The object \object{BL\,2007+777} has $V$=16.5 and $z$=0.342 
(Stickel \& K\"uhr
 \cite{sti96}).  
This radio source has been studied with different techniques
and at different frequencies in the last 20 years.  
Its kpc scale has been studied, e.g., by Antonucci \et\ (\cite{ant86}), Kollgaard \et\ 
(\cite{kol92}), and Murphy \et\ (\cite{mur93}).
Those authors presented maps with compact emission, 
and extended components at 8.5 and 15.8\,arcsec 
to the west, and another {extended component} 
at 11\,arcsec to the east, of
the brightest {compact} component. 
At pc scales, the radio source is also bright.  
All VLBI maps (Eckart \et\ \cite{eck87},
Pearson \& Readhead \cite{pea88}, Guirado \et\ \cite{gui95a}, 
Xu \et\ \cite{xu95}, Fey \& Charlot \cite{fey97}, Guirado \et\ \cite{gui98}, 
Ros \et\ \cite{ros99}) show emission west of the main component
with the brightest features
about 5--6\,mas of the core, and emission extending {out} to 20\,mas at 
5\,GHz for epoch 1992.47 (Xu \et\ \cite{xu95}) and {out}
to 40\,mas at 2.3\,GHz for epoch 1995.77 (Fey \& Charlot \cite{fey97}).
\object{BL\,2007+777} displays strong flux variability.  
Data from UMRAO at 8\,GHz show that the source changed the total flux density
from $\sim$0.8\,Jy at 1996.5 to $\sim$1.8\,Jy at 1997.2.  
Throughout 1997
the source maintained a total
flux density of $\sim$1.7\,Jy at 8.4\,GHz, decreasing to
$\sim$1.4\,Jy for our second epoch {in mid 1999}.

%
\begin{figure}[htbp]
\vspace*{225pt}
\includegraphics{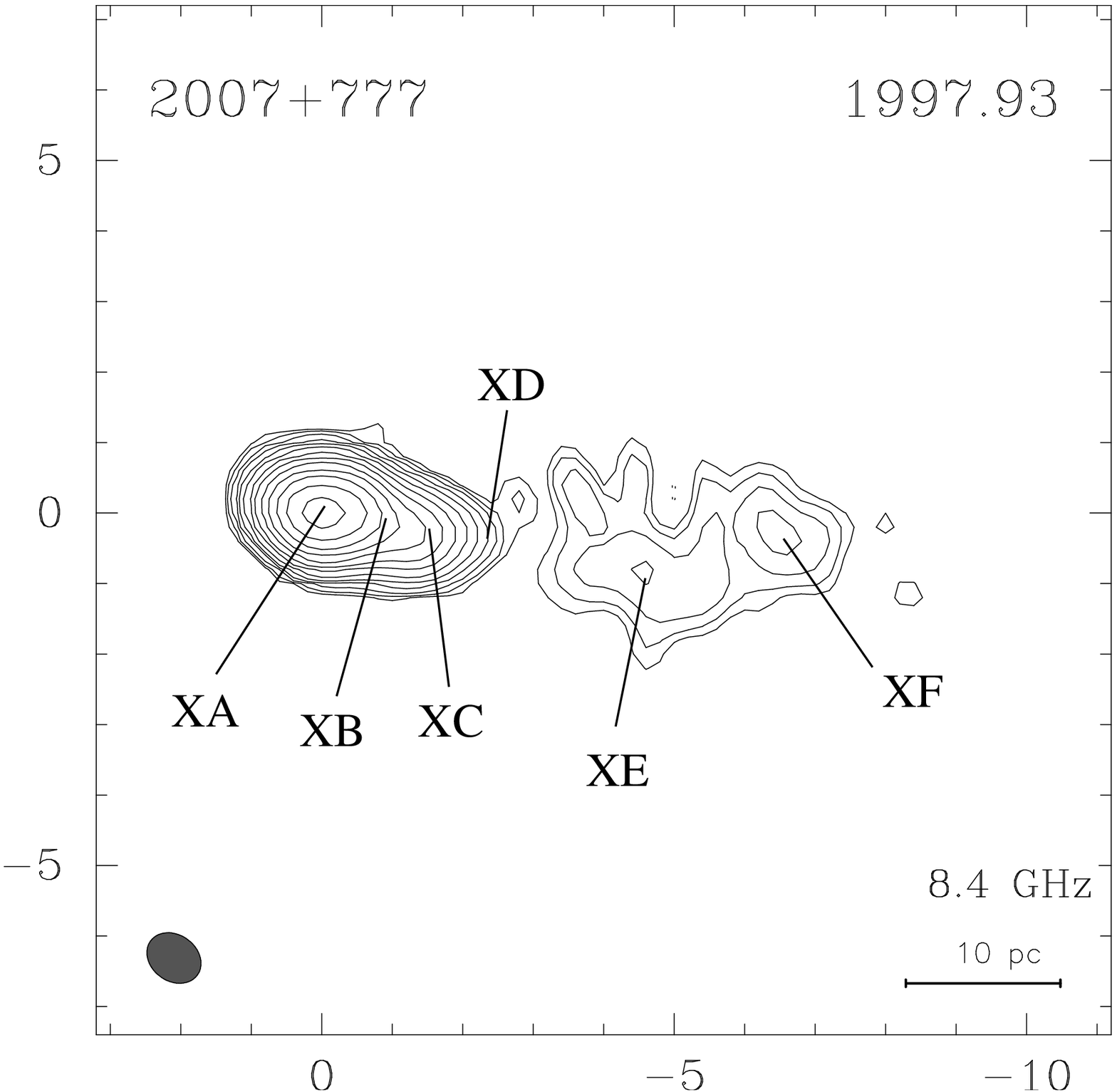}
\vspace*{225pt}
\includegraphics{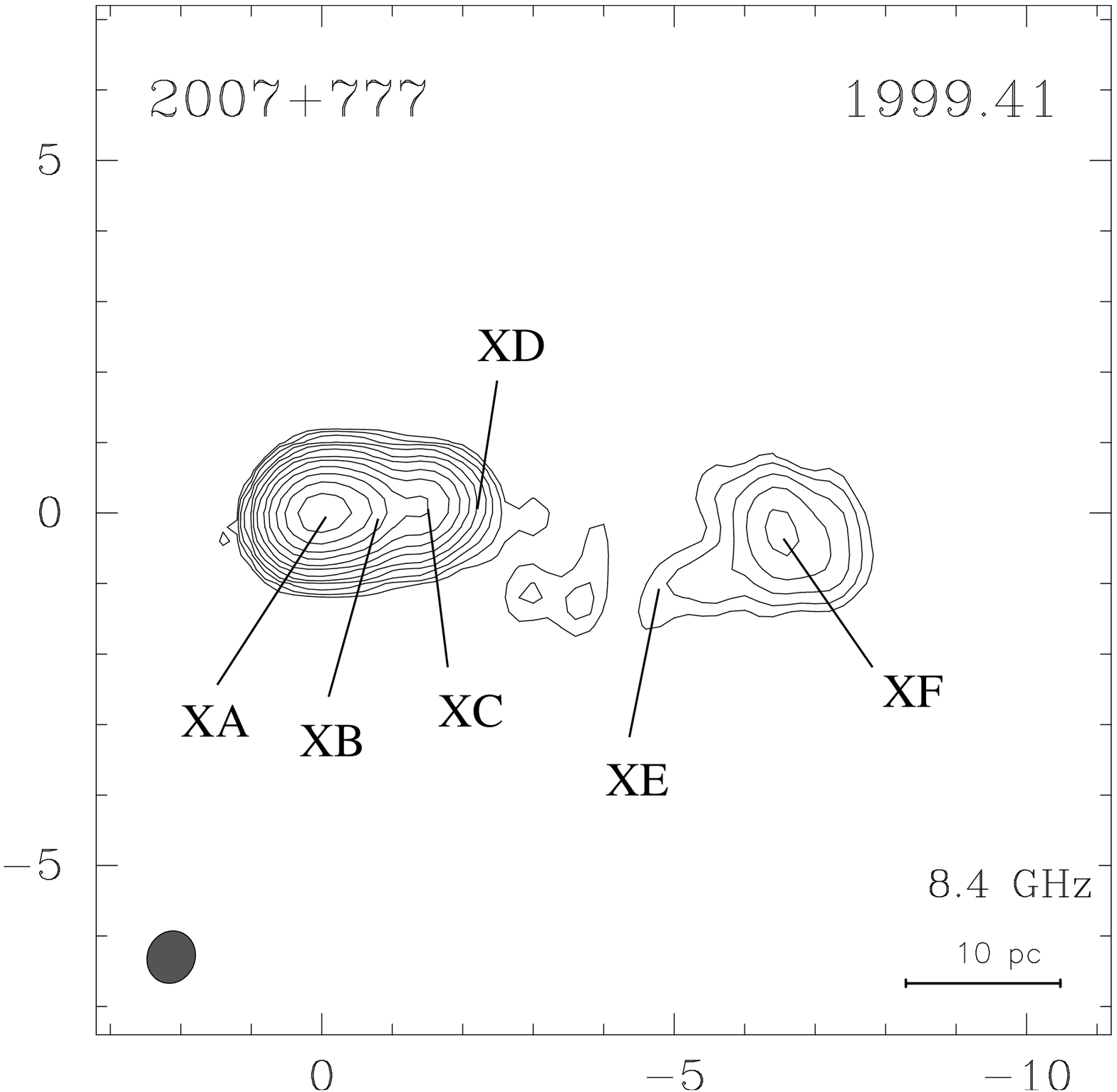}
\caption{VLBA images of \object{BL\,2007+777}, observed on 6 December 1997
(1997.93) and 28 May 1999 (1999.41).  
Axes are relative $\alpha$ and $\delta$ in mas.
See Table 1 for contour levels, 
synthesized beam sizes (bottom left in the maps),
and peak flux densities.
{See Table~\ref{table:modelfit} for component parametrization.}}
\label{fig:map2007}
\end{figure}

Our maps (Fig.\ \ref{fig:map2007}) 
recover flux densities
of 1.305 and 0.969\,Jy for the first and second epochs, respectively.
{Changes} in the structure are evident.  
A secondary component about 2\,mas (labeled XC+XD in
Fig.\ \ref{fig:map2007}) west of the brightest feature (XA) 
{appears at a different P.A., having moved}
northwards from the first to the second epoch.
This is confirmed by the 5\,GHz VLBI Space Orbiting Programme data from
{epochs close to ours (Krichbaum \et\ 2000)}.
A model fit with Gaussian components (Table \ref{table:modelfit})
shows two components for the $\sim$2\,mas feature, 
XC and XD, that change the P.A.\
from $-102$\degr\ and $-98$\degr\ in the first epoch to $-88$\degr 
and $-86$\degr, respectively.  
Their distances from
the brightest feature change from 1.2
and 1.6\,mas to 1.4 and 1.7\,mas, respectively. 
Assuming that component XA is stationary,
these position changes correspond to {northward superluminal motion
(not only from the core outwards)}.
The {core} area is double, with XA with similar flux densities
in both epochs, and XB decaying to a 30\% of its previous flux
densities in 1999.41.
Guirado \et\ (\cite{gui00}) report a changing double structure in
this core area at 43\,GHz, which questions any registration 
of this region without astrometric information.

There are also structural changes in the components labeled
XE and XF.  While XE decays strongly in emission, XF becomes more prominent
(it doubles its flux density from the first to the second epoch).
This is consistent with space VLBI observations from 
Krichbaum \et\ (2000), which report changes in the blobs of 
\object{BL\,2007+777} within a month in the region 6--7\,mas from the ``core".  
The {forthcoming}
exact alignment of the features provided by the astrometry should
constrain the proper motions of the components of the jet of this 
radio source.

\section{Summary\label{sec:summary}}

We present 8.4\,GHz VLBA maps 
of the thirteen extragalactic radio sources from the S5 polar cap sample.
For two epochs, all sources of the sample have been imaged
(Figs.\ \ref{fig:map0016}-\ref{fig:map2007}) 
with mas-resolution to the mJy level (Table~\ref{table:results-mapping}).
In all cases, we have modeled
the interferometric visibilities with elliptical Gaussian models 
(Table~\ref{table:modelfit}).

We can report some morphological changes
for selected radio sources.  The
sources with more evident variations are \object{QSO\,0016+731},
\object{QSO\,0836+710}, \object{QSO\,1928+738} and \object{BL\,2007+777}.
For \object{QSO\,0016+731} the variations have been surprising, since the 
source is rather compact and does not show a prominent jet structure
larger than 2--3\,mas at 8.4\,GHz.  Its change in morphology 
is associated to the strong decrease in
flux density of one of the components
from the first to the second epoch.
\object{QSO\,0836+710} and \object{QSO\,1928+738} show the familiar
changes  in structure produced by emerging components in
a core-jet system, as reported by Otterbein \et\ (\cite{ott98})
for the former and Hummel \et\ (\cite{hum92a}) for the latter.
The BL\,Lac object 2007+777 displays an interesting, and
hard to explain change
in position angle towards north for components near to the core,
fact also reported
by Krichbaum \et\ (\cite{kri00}) from 5\,GHz VSOP observations.

The images are the result of first and second epochs
of a phase-delay astrometric program intended to check the absolute
kinematics of all radio sources of the sample with precisions
better than 100\,$\mu$as.  Such accurate results
will be obtained after the mapping and astrometric reduction
of more observing epochs is carried out.
This multi-source astrometric approach provides a large
number of constraints for all the relative source pairs, which
allow a precise registration of the maps through the observing
epochs (as shown by Ros \et\ (\cite{ros99}) for triplets of
radio sources).

We have recently extended our astrometric programme to 15 and
43\,GHz.  At these frequencies, and based on test observations,
we expect to attain astrometric precisions of 50 and
20\,$\mu$as, respectively.  The two-fold and five-fold improved
resolution of those observations with respect to the observations
presented here, combined with the expected astrometric precisions,
will allow precise registrations of all the sources at each
and all wavelengths for all epochs and will provide unprecedented
spectral information of components of milliarcsecond sources.
The determination of the detailed kinematics and spectral 
content of the compact components of a complete sample of 
radio sources should then turn out to be a decisive element in
our understanding of the activity around the cores of these compact
radio sources and in a definitive test of the standard jet model.

\begin{acknowledgements}
This work has been partially financed
by Grant PB96-0782 of the Spanish DGICYT.
This research has made use of data from the University of
Michigan Radio Astronomy Observatory which is supported 
by the National Science Foundation and by funds from the University 
of Michigan.  NRAO is operated under license by Associated Universities Inc.,
under cooperative agreement with NSF.
\end{acknowledgements}

\end{document}